\documentclass[usenatbib]{mnras}
\usepackage[T1]{fontenc}
\DeclareRobustCommand{\VAN}[3]{#2}
\let\VANthebibliography\thebibliography
\def\thebibliography{\DeclareRobustCommand{\VAN}[3]{##3}\VANthebibliography}
\usepackage{graphicx}	
\usepackage{amsmath}	
\usepackage{amssymb}	
\usepackage{color}
\usepackage{url}
\usepackage{dsfont}
\usepackage{subfig}
\usepackage{CJKutf8}
\usepackage{threeparttable}

\newcommand\lsim{\mathrel{\rlap{\lower4pt\hbox{\hskip1pt$\sim$}}
        \raise1pt\hbox{$<$}}}
\newcommand\gsim{\mathrel{\rlap{\lower4pt\hbox{\hskip1pt$\sim$}}
        \raise1pt\hbox{$>$}}}
\newcommand{\lya}{\ifmmode\mathrm{Ly}\alpha\else{}Ly$\alpha$\fi}
\newcommand{\lyb}{\ifmmode\mathrm{Ly}\beta\else{}Ly$\beta$\fi}
\newcommand{\igm}{\ifmmode\mathrm{IGM}\else{}IGM\fi}
\newcommand{\lae}{\ifmmode\mathrm{LAE}\else{}LAE\fi}
\newcommand{\h}{\ifmmode\mathrm{H}\else{}H\fi}
\newcommand{\hi}{\ifmmode\mathrm{H\,{\scriptscriptstyle I}}\else{}H\,{\scriptsize I}\fi}
\newcommand{\hii}{\ifmmode\mathrm{H\,{\scriptscriptstyle II}}\else{}H\,{\scriptsize II}\fi}
\newcommand{\cmb}{\ifmmode\mathrm{CMB}\else{}CMB\fi}
\newcommand{\qso}{\ifmmode\mathrm{QSO}\else{}QSO\fi}
\newcommand{\eor}{\ifmmode\mathrm{EoR}\else{}EoR\fi}
\newcommand{\heii}{\ifmmode\mathrm{He\,{\scriptscriptstyle II}}\else{}He\,{\scriptsize II}\fi}
\newcommand{\heiii}{\ifmmode\mathrm{He\,{\scriptscriptstyle III}}\else{}He\,{\scriptsize III}\fi}
\newcommand{\ciii}{\ifmmode\mathrm{C\,{\scriptscriptstyle III]}}\else{}C\,{\scriptsize III]}\fi}
\newcommand{\oiii}{\ifmmode\mathrm{O\,{\scriptscriptstyle III}}\else{}O\,{\scriptsize III}\fi}
\newcommand{\aliii}{\ifmmode\mathrm{Al\,{\scriptscriptstyle III}}\else{}Al\,{\scriptsize III}\fi}
\newcommand{\mgii}{\ifmmode\mathrm{Mg\,{\scriptscriptstyle II}}\else{}Mg\,{\scriptsize II}\fi}
\newcommand{\fe}{\ifmmode\mathrm{Fe}\else{}Fe\fi}
\newcommand{\nv}{\ifmmode\mathrm{N\,{\scriptscriptstyle V}}\else{}N\,{\scriptsize V}\fi}
\newcommand{\niv}{\ifmmode\mathrm{N\,{\scriptscriptstyle IV]}}\else{}N\,{\scriptsize IV]}\fi}
\newcommand{\cii}{\ifmmode\mathrm{C\,{\scriptscriptstyle II}}\else{}C\,{\scriptsize II}\fi}
\newcommand{\civ}{\ifmmode\mathrm{C\,{\scriptscriptstyle IV}}\else{}C\,{\scriptsize IV}\fi}
\newcommand{\siv}{\ifmmode\mathrm{Si\,{\scriptscriptstyle IV}}\else{}Si\,{\scriptsize IV}\fi}
\newcommand{\siii}{\ifmmode\mathrm{Si\,{\scriptscriptstyle II}}\else{}Si\,{\scriptsize II}\fi}
\newcommand{\siiii}{\ifmmode\mathrm{Si\,{\scriptscriptstyle III]}}\else{}Si\,{\scriptsize III]}\fi}
\newcommand{\ovi}{\ifmmode\mathrm{O\,{\scriptscriptstyle VI}}\else{}O\,{\scriptsize VI}\fi}
\newcommand{\sioiv}{\ifmmode\mathrm{Si\,{\scriptscriptstyle IV}\,\plus O\,{\scriptscriptstyle IV]}}\else{}Si\,{\scriptsize IV}\,+O\,{\scriptsize IV]}\fi}

\newcommand{\cmmc}{\textsc{\small 21CMMC}}
\newcommand{\cmfst}{\textsc{\small 21CMFAST}}

\usepackage{newtxtext,newtxmath}

\pdfoutput=1

\title[EoR parameter inference using the 2DPS]{Inferring astrophysical parameters using the 2D cylindrical power spectrum from reionisation}

\author[B. Greig et al.]{
Bradley Greig$^{1,2,3}$\thanks{E-mail:~bradley.greig@anu.edu.au}, David Prelogovi\'{c}$^{4}$, Yuxiang Qin$^{2,3}$, Yuan-Sen Ting (丁源森)$^{1,5,6,7,8}$ \newauthor \& Andrei Mesinger$^{4}$ \\
$^1$Research School of Astronomy \& Astrophysics, Australian National University, Canberra, ACT 2611, Australia \\
$^2$School of Physics, University of Melbourne, Parkville, VIC 3010, Australia \\
$^3$ARC Centre of Excellence for All-Sky Astrophysics in 3 Dimensions (ASTRO 3D) \\
$^4$Scuola Normale Superiore, Piazza dei Cavalieri 7, I-56125 Pisa, Italy \\
$^5$School of Computing, Australian National University, Acton ACT 2601, Australia \\
$^6$Department of Astronomy, The Ohio State University, Columbus, OH 45701, USA \\
$^7$Center for Cosmology and AstroParticle Physics (CCAPP), The Ohio State University, Columbus, OH 43210, USA \\
$^8$Department of Physics, Faculty of Science, Universiti Malaya, 50603 Kuala Lumpur, Malaysia \\
}

\pubyear{2024}

\begin{document}
\label{firstpage}
\pagerange{\pageref{firstpage}--\pageref{lastpage}}
\begin{CJK}{UTF8}{gkai} 
\maketitle
\end{CJK}

\begin{abstract}
Enlightening our understanding of the first galaxies responsible for driving reionisation requires detecting the 21-cm signal from neutral hydrogen. Interpreting the wealth of information embedded in this signal requires Bayesian inference. Parameter inference from the 21-cm signal is primarily restricted to the spherically averaged power spectrum (1D PS) owing to its relatively straightforward derivation of an analytic likelihood function enabling traditional Monte-Carlo Markov-Chain (MCMC) approaches. However, in recent years, simulation-based inference (SBI) has become feasible which removes the necessity of having an analytic likelihood, enabling more complex summary statistics of the 21-cm signal to be used for Bayesian inference. In this work, we use SBI, specifically marginal neural ratio estimation to learn the likelihood-to-evidence ratio with \textsc{Swyft}, to explore parameter inference using the cylindrically averaged 2D PS. Since the 21-cm signal is anisotropic, the 2D PS should yield more constraining information compared to the 1D PS which isotropically averages the signal. For this, we consider a mock 1000 hr observation of the 21-cm signal using the SKA and compare the performance of the 2D PS relative to the 1D PS. Additionally, we explore two separate foreground mitigation strategies, perfect foreground removal and wedge avoidance. We find the 2D PS outperforms the 1D PS by improving the marginalised uncertainties on individual astrophysical parameters by up to $\sim30-40$ per cent irrespective of the foreground mitigation strategy. Primarily, these improvements stem from how the 2D PS distinguishes between the transverse, $k_{\perp}$, and redshift dependent, $k_{\parallel}$ information which enables greater sensitivity to the complex reionisation morphology.  
\end{abstract}




\section{Introduction}

Roughly 400,000 years after the Big Bang recombination occurs, whereby the photons and baryons of the primordial plasma decouple and the baryons subsequently combine to form neutral hydrogen. After this point the omnipresence of neutral hydrogen enshrouds the Universe in a fog rendering it opaque to most forms of radiation. This fog persists until the ignition of the first star-formation episodes within the primordial galaxies, referred to as the cosmic dawn (CD), which emit copious amounts of ultra-violet (UV) photons into the intergalactic medium (IGM) and ionise their local neighbourhood. Over time, as these galaxies grow and become more abundant their cumulative UV output accelerates the eradication of this fog, rendering the IGM essentially completely ionised. This process, is referred to as the Epoch of Reionisation (EoR).

Directly observing these primordial galaxies is nigh on impossible. Their vast distance from us makes them extremely faint, and this is prior to the extinction of their radiation by the neutral IGM. All however is not lost. The primordial galaxies imprint their signal on the neutral IGM, enabling us to indirectly infer their presence by measuring the 21-cm hyperfine spin-flip transition of the neutral hydrogen. This signal is observable by detecting the differential intensity of radiation emitted by the neutral hydrogen relative to a uniform background source, for example, the Cosmic Microwave Background \citep[see e.g.][]{Gnedin:1997p4494,Madau:1997p4479,Shaver:1999p4549,Tozzi:2000p4510,Gnedin:2004p4481,Furlanetto:2006p209,Morales:2010p1274,Pritchard:2012p2958}. As this frequency (redshift) dependent signal originates from the IGM, detecting it yields a three-dimensional, time-evolving picture of the thermal and ionisation state of the IGM throughout reionisation.

Accessing the wealth of information embedded in the 3D cosmic 21-cm signal requires large-scale radio interferometer experiments to pick up the spatially varying signal. Specifically, we measure the complex visibilities of the interference fringes from the arriving signal which are naturally represented by a Fourier transform. This signal can then be split into two components, $k_{\parallel}$ which corresponds to the line-of-sight (frequency) dependent aspect of the signal and $k_{\perp}$ which describes the 2D spatial variation of the signal across the sky. Typically, given the cosmic 21-cm signal is incredibly faint relative to the bright foregrounds, in order to boost its signal to noise we compress the available information by measuring the 1D spherically averaged power spectrum (PS). This characterises the statistical properties of the 21-cm signal by describing the variance as a function of spatial scale. It is this quantity that is sought by the first generation interferometer experiments such as the Low-Frequency Array (LOFAR; \citealt{vanHaarlem:2013p200}), the Murchison Wide Field Array (MWA; \citealt{Tingay:2013p2997,Wayth:2018}), the Precision Array for Probing the Epoch of Reionisation (PAPER; \citealt{Parsons:2010p3000}), the Owens Valley Radio Observatory Long Wavelength Array (OVRO-LWA; \citealt{Eastwood:2019}) and the upgraded Giant Metrewave Radio Telescope (uGMRT; \citealt{Gupta:2017}).

Importantly, the 21-cm signal is non-Gaussian owing to the complex 3D morphology of the ionised regions. Therefore computing the 1D PS performs sub-optimal compression as we disregard valuable non-Gaussian information. The lower sensitivity of these first generation of experiments necessitates the trade-off in information loss in order to boost the overall signal to noise with the 1D PS. However, for the forthcoming Square Kilometre Array (SKA; \citealt{Mellema:2013p2975,Koopmans:2015}) this should no longer be a concern with the increased sensitivity theoretically enabling more complex summary statistics of the 21-cm signal beyond the 1D PS. In fact, the SKA has been specifically designed with tomographic imaging in mind, opening up a wealth of possibilities for analysing the 21-cm signal.

This has lead to the explosion of studies exploring alternative probes of the 21-cm signal beyond the 1D PS. For example, with the bispectrum \citep[e.g.][]{Yoshiura:2015,Shimabukuro:2016,Majumdar:2018,Watkinson:2019,Hutter:2020,Majumdar:2020,Kamran:2021}, position dependent power spectrum \citep{Giri:2019b}, one-point statistics \citep{Watkinson:2014,Shimabukuro:2015,Kubota:2016,Banet:2021,Gorce:2021}, morphological and topographical coefficients extracted from 21-cm images \citep[e.g.][]{Yoshiura:2017,Bag:2019,Chen:2019,Elbers:2019,Kapahtia:2019,Gazagnes:2021,Giri:2021,Kapahtia:2021}, the ionised bubble size distribution \citep{Kakiichi:2017,Giri:2018a,Giri:2018b,Giri:2019a,Bianco:2021} and the compression of 21-cm images using the wavelet scattering transform \citep[e.g.][]{Greig:2022,Hothi:2023}.

Importantly, in order to extract the astrophysical properties of these first galaxies we must perform Bayesian inference. Typically, this demands generating 3D reionisation simulations on-the-fly within a Monte-Carlo Markov-Chain (MCMC) framework (e.g. \cmmc{} \citealt{Greig:2015p3675,Greig:2017p8496,Greig:2018,Park:2019}) to compare against an observation of the 21-cm signal. However, this approach is extremely restrictive as it requires defining an analytic expression to compute the likelihood of the 21-cm signal given the model astrophysical parameter set. As a result, essentially none of the aforementioned alternative statistics to the 1D PS have been explored rigorously within a Bayesian inference context. Instead, most resort to the Fisher information matrix \citep{Fisher:1935}, which still imposes an implicit Gaussian assumption, to provide simple forecasts \citep[e.g.][]{Shimabukuro:2017,Greig:2022,Hothi:2023} or adopt several simplifying assumptions regarding the covariance and likelihood form \citep{Watkinson:2021,Tiwari:2021}. Alternatively, instead of performing direct inference one can apply regression in an attempt to extract astrophysical information from the 21-cm signal by bypassing the need for a summary statistic entirely through the application of convolutional neural networks (CNNs) trained directly on 2D or 3D images of the 21-cm signal to extract astrophysical information \citep[e.g.][]{Gillet:2019,Hassan:2019,LaPlante:2019,Hassan:2020,Kwon:2020,Mangena:2020,Prelogovic:2021}.

However, we can completely forego this restrictive requirement of an analytic likelihood through the concept of likelihood-free or simulation-based inference (SBI; see e.g. \citealt{Cranmer:2020} for a recent review). Essentially, we apply machine learning on a training set of simulated data to learn our likelihood function (neural likelihood estimation; NLE) or the likelihood-to-evidence ratio (neural ratio estimation; NRE) after which we can perform an MCMC to obtain our posteriors or bypass the likelihood estimation entirely to direclty obtain our posterior distribution given our data (neural posterior estimation; NPE). The power of such an approach is that it enables us to explore any complex or non-Gaussian summary statistic or feature extraction method applied to the 21-cm signal. Our only requirement is the generation of the simulated training set. Realising this potential, in recent years SBI has been gaining traction for tackling astrophysical inference from the cosmic 21-cm signal \citep[e.g.][Greig et al. in prep]{Zhao:2022,Zhao:2022b,Prelogovic:2023,Saxena:2023}.

In this work we choose to explore the oft overlooked 2D cylindrically averaged (2D PS) for astrophysical parameter inference\footnote{\citet{Mondal:2022} performed an initial exploratory analysis of the multi-frequency angular power spectrum (MAPS), which has some analogies to the 2D PS. For a basic three parameter astrophysical model the MAPS was found to outperform the 1D PS.}. Its previous omission stemmed from our inability to define a robust likelihood function along with the additional computational overheads required for estimating the 21-cm signal covariance. Specifically, we follow the approach of our companion work (Greig et al. in prep) and perform our SBI using Marginal Neutral Ratio Estimation (MNRE; \citealt{Miller:2021}) using the publicly available \textsc{\small Python} package, \textsc{\small Swyft}\footnote{https://github.com/undark-lab/swyft} \citep{Miller:2022} whose goal is to learn the marginal likelihood-to-evidence ratios for each individual parameter. Rather than spherically averaging over the $k_{\parallel}$ and $k_{\perp}$ components of the signal into a single $k$ and losing morphological information, the 2D PS keeps these components distinct, better separating out the 2D structural information from the spatial fluctuations across the sky ($k_{\perp}$) from the frequency varying component along the line-of-sight, $k_{\parallel}$. Further, the 2D PS more naturally follows the signal characteristics obtained from radio interferometry allowing us to more cleanly deal with foreground contamination (e.g. the `wedge'). Although it is still a Gaussian statistic, and therefore still sub-optimal, it should lose less information than the 1D PS. In fact, in a complimentary study by \citet{Prelogovic:2024} exploring the information content of a variety of 21-cm summaries using Fisher Matrices these authors predict improvements in the variance on the astrophysical parameters of $\sim15$ per cent. Further, the 2D PS is considerably more straightforward to measure observationally and should require less integration time to achieve sufficient sensitivity in comparison to many of the aforementioned non-Gaussian approaches. Therefore it is an important and valuable summary statistic to explore.

The remainder of this paper is organised as follows. In Section~\ref{sec:21cm} we summarise our 21-cm simulations using \cmfst{} and in Section~\ref{sec:setup} we describe our SBI setup with \textsc{Swyft} including the generation of our database of 21-cm simulations and our mock observation. In Section~\ref{sec:results} we then perform our comparison of the 2D PS to the 1D PS for different foreground mitigation strategies before concluding with our closing remarks in Section~\ref{sec:conclusion}. Unless stated otherwise, all quantities are in in co-moving units and we adopt the cosmological parameters:  ($\Omega_\Lambda$, $\Omega_{\rm M}$, $\Omega_b$, $n$, $\sigma_8$, $H_0$) = (0.69, 0.31, 0.048, 0.97, 0.81, 68 km s$^{-1}$ Mpc$^{-1}$), consistent with recent results from the Planck mission \citep{Planck:2020}.

\section{Simulating the 21-cm signal} \label{sec:21cm}

To simulate the 3D cosmic 21-cm signal emanating from the neutral hydrogen during reionisation we use the semi-numerical simulation code \cmfst{}\footnote{https://github.com/21cmfast/21cmFAST}\citep{Mesinger:2007p122,Mesinger:2011p1123}. In particular, we use the latest public release, v3 \citep{Murray:2020}, and adopt the \citet{Park:2019} flexible galaxy parameterisation to describe the UV and X-ray properties of the galaxy population. In this section we outline the main ingredients of \cmfst{}, in particular focussing on the astrophysical parameters within the model we seek to constrain using parameter inference. For additional details and discussions we defer the reader to these earlier publications.

\subsection{Galaxy UV properties}

First, it is assumed that the stellar mass, $M_{\ast}$, of a galaxy depends on its host halo mass, $M_{\rm h}$ \citep[e.g.][]{Kuhlen:2012p1506,Dayal:2014b,Behroozi:2015p1,Mitra:2015,Mutch:2016,Ocvirk:2016,Sun:2016p8225,Yue:2016,Hutter:2020} via the following relation:
\begin{eqnarray} 
M_{\ast}(M_{\rm h}) = f_{\ast}\left(\frac{\Omega_{\rm b}}{\Omega_{\rm m}}\right)M_{\rm h},
\end{eqnarray}
with $f_{\ast}$ being the fraction of galactic gas in stars and $\Omega_{\rm b}$ and $\Omega_{\rm m}$ being the baryonic and total matter content of the Universe. $f_{\ast}$ also depends on its host halo mass,
\begin{eqnarray}
f_{\ast} = f_{\ast, 10}\left(\frac{M_{\rm h}}{10^{10}\,M_{\odot}}\right)^{\alpha_{\ast}}.
\end{eqnarray}
dependent on the two free parameters, $\alpha_{\ast}$ and its normalisation, $f_{\ast, 10}$, for a dark matter halo mass of $10^{10}$~$M_{\odot}$. This power-law behaviour directly follows from semi-empirical fits to observations \citep[e.g.][]{Harikane:2016,Tacchella:2018,Behroozi:2019,Stefanon:2021} and semi-analytic model predictions \citep[e.g][]{Mutch:2016,Yung:2019,Hutter:2020}.

The stellar mass is then converted into a star-formation rate (SFR) by dividing by a characteristic time-scale, $t_{\ast}$, which is a free parameter of the model and is defined to be a fraction, $t_{\ast}$ $\in[0.05,1]$, of the Hubble time, $H^{-1}(z)$:
\begin{eqnarray} \label{eq:sfr}
\dot{M}_{\ast}(M_{\rm h},z) = \frac{M_{\ast}}{t_{\ast}H^{-1}(z)}.
\end{eqnarray}

Similarly as above, the fraction of UV photons that escape their host galaxy and enter into the IGM, $f_{\rm esc}$, also depend on their host halo mass,
\begin{eqnarray} \label{}
f_{\rm esc} = f_{\rm esc, 10}\left(\frac{M_{\rm h}}{10^{10}\,M_{\odot}}\right)^{\alpha_{\rm esc}},
\end{eqnarray}
giving rise to an additional two free parameters, $\alpha_{\rm esc}$ and $f_{\rm esc, 10}$.

Not all dark matter haloes can contribute to reionisation. Internal feedback mechanisms and/or inefficient gas cooling can suppress star-formation in low mass haloes. This behaviour is parameterised via an effective duty-cycle:
\begin{eqnarray} \label{eq:duty}
f_{\rm duty} = {\rm exp}\left(-\frac{M_{\rm turn}}{M_{\rm h}}\right).
\end{eqnarray}
with $(1 - f_{\rm duty})$ defining the fraction of star-forming galaxies that are suppressed below a characteristic mass scale $M_{\rm turn}$ \citep[e.g.][]{Shapiro:1994,Giroux:1994,Hui:1997,Barkana:2001p1634,Springel:2003p2176,Mesinger:2008,Okamoto:2008p2183,Sobacchi:2013p2189,Sobacchi:2013p2190}.

\subsection{Galaxy X-ray properties}

In addition to contributing the UV photons responsible for driving reionisation, the first galaxies also emit X-ray photons which escape and heat the cold IGM gas. The origin of the X-ray photons is thought to be stellar remnants left over from earlier star-formation episodes. To model the X-ray heating caused by these energetic photons \cmfst{} computes a cell-by-cell angle-averaged specific X-ray intensity, $J(\boldsymbol{x}, E, z)$, (in erg s$^{-1}$ keV$^{-1}$ cm$^{-2}$ sr$^{-1}$), 
\begin{equation} \label{eq:Jave}
J(\boldsymbol{x}, E, z) = \frac{(1+z)^3}{4\pi} \int_{z}^{\infty} dz' \frac{c dt}{dz'} \epsilon_{\rm X}  e^{-\tau}.
\end{equation}
by integrating the co-moving X-ray specific emissivity, $\epsilon_{\rm X}(\boldsymbol{x}, E_e, z')$ back along the light-cone accounting for IGM attenuation, $e^{-\tau}$. The specific emitted emissivity, $E_{\rm e} = E(1 + z')/(1 + z)$, is then,
\begin{equation} \label{eq:emissivity}
\epsilon_{\rm X}(\boldsymbol{x}, E_{\rm e}, z') = \frac{L_{\rm X}}{\rm SFR} \left[ (1+\bar{\delta}_{\rm nl}) \int^{\infty}_{0}{\rm d}M_{\rm h} \frac{{\rm d}n}{{\rm d}M_{\rm h}}f_{\rm duty} \dot{M}_{\ast}\right],
\end{equation}
where $\bar{\delta}_{\rm nl}$ is the mean, non-linear overdensity in a shell centred on the simulation cell $(\boldsymbol{x}, z)$ and the quantity in square brackets is the SFR density along the light-cone with $\frac{{\rm d}n}{{\rm d}M_{\rm h}}$ corresponding to the halo mass function (HMF)\footnote{Throughout this work we adopt the Sheth-Tormen HMF \citep{Sheth:2001} as our fiducial HMF.}. The quantity $L_{\rm X}/{\rm SFR}$ (erg s$^{-1}$ keV$^{-1}$ $M^{-1}_{\odot}$ yr) is the specific X-ray luminosity per unit star formation escaping the host galaxies which depends on the spectral energy distribution describing the source of X-rays, $L_{\rm X} \propto E^{- \alpha_X}$. Throughout, we adopt $\alpha_{\rm X} = 1$, consistent with local Universe observations of high-mass X-ray binaries  \citep[e.g.][]{Mineo:2012p6282,Fragos:2013p6529,Pacucci:2014p4323}.

Finally, we normalise $L_{\rm X}/{\rm SFR}$ by its integrated soft-band ($<2$~keV) luminosity per SFR (in erg s$^{-1}$ $M^{-1}_{\odot}$ yr),
\begin{equation} \label{eq:normL}
  L_{{\rm X}<2\,{\rm keV}}/{\rm SFR} = \int^{2\,{\rm keV}}_{E_{0}} dE_e ~ L_{\rm X}/{\rm SFR}. 
\end{equation}
with $E_0$ denoting the minimum X-ray photon energy capable of escaping the host galaxy into the IGM.

\subsection{Ionisation and Thermal State of the IGM}

The thermal state of the IGM is computed via the IGM spin temperature, $T_{\rm S}$, which is determined by self-consistently computing the heating and ionisation rates owing to structure formation, Compton scattering off CMB photons, heating following partial ionisations as well as X-ray heating and ionisations. To calculate $T_{\rm S}$ we determine its weighted mean,
\begin{eqnarray}
T^{-1}_{\rm S} = \frac{T^{-1}_{\rm CMB} + x_{\alpha}T^{-1}_{\alpha} + x_{\rm c}T^{-1}_{\rm K}}{1 + x_{\alpha} + x_{\rm c}},
\end{eqnarray}
where $T_{\rm K}$, $T_{\alpha}$ and $T_{\rm CMB}$ are the gas, Lyman-$\alpha$ (Ly$\alpha$) colour and CMB temperatures. $T_{S}$ depends on the local gas density and \lya{} radiation intensity, with the \lya{} background sourced by the cumulative sum of X-ray excitations of neutral hydrogen atoms and direct stellar emission of Lyman band photons by the first galaxies. The quantities $x_{\alpha}$ and $x_{\rm c}$ are the coupling coefficients for the Wouthuysen-Field mechanism \citep{Wouthuysen:1952p4321,Field:1958p1} and between the free elections and CMB photons, respectively.

Calculating the 3D ionisation of the IGM requires the application of excursion-set theory \citep{Furlanetto:2004p123} on the evolved density field. This compares the cumulative number of ionising photons, $n_{\rm ion}$, to the total number of neutral hydrogen atoms plus cumulative recombinations, $\bar{n}_{\rm rec}$ \citep{Sobacchi:2014p1157} within spheres of decreasing radii, $R$, and corresponding overdensity, $\delta_{R}$. Evaluated within each individual simulation voxel, a voxel is deemed to be ionised when
\begin{eqnarray} \label{eq:ioncrit}
\bar{n}_{\rm ion}(\boldsymbol{x}, z | R, \delta_{R}) \geq (1 + \bar{n}_{\rm rec})(1-\bar{x}_{e}),
\end{eqnarray}
where the $(1-\bar{x}_{e})$ factor includes the contribution of X-rays to ionisations and,
\begin{eqnarray} \label{eq:ioncrit2}
n_{\rm ion} = \bar{\rho}^{-1}_b\int^{\infty}_{0}{\rm d}M_{\rm h} \frac{{\rm d}n(M_{h}, z | R, \delta_{R})}{{\rm d}M_{\rm h}}f_{\rm duty} \dot{M}_{\ast}f_{\rm esc}N_{\gamma/b}.
\end{eqnarray}
Here, $\bar{\rho}_b$ is the mean baryon density and $N_{\gamma/b}$ is the total number of ionising photons produced per stellar baryon\footnote{By default this is assumed to be $N_{\gamma/b}=5000$ consistent with a Salpeter initial mass function \citep{Salpeter:1955}}.

\subsection{21-cm Brightness Temperature}

The quantity we measure observationally is the brightness temperature, $\delta T_{\rm b}(\nu)$, the differential intensity of the neutral hydrogen illuminated by the CMB \citep{Furlanetto:2006p209},
\begin{eqnarray} \label{eq:21cmTb}
\delta T_{\rm b}(\nu) &=& \frac{T_{\rm S} - T_{\rm CMB}(z)}{1+z}\left(1 - {\rm e}^{-\tau_{\nu_{0}}}\right)~{\rm mK},
\end{eqnarray}
and
\begin{eqnarray}
\tau_{\nu_{0}} &\propto& (1+\delta_{\rm nl})(1+z)^{3/2}\frac{x_{\hi{}}}{T_{\rm S}}\left(\frac{H}{{\rm d}v_{\rm r}/{\rm d}r+H}\right),
\end{eqnarray}
where $\tau_{\nu_{0}}$, is the optical depth of the neutral hydrogen which depends on the local gas overdensity, $\delta_{\rm nl} \equiv \rho/\bar{\rho} - 1$, the neutral hydrogen fraction, $x_{\hi{}}$, the Hubble parameter, $H(z)$, and the line-of-sight gradient of the peculiar velocity. For simplicity the spatial dependence of the quantities have been omitted and it is evaluated at the redshift $z = \nu_{0}/\nu - 1$.

\section{Simulation Based Inference Setup} \label{sec:setup}

\subsection{Parameter inference with \textsc{\small Swyft}}

In parameter inference, the quantity of interest is the posterior, $p(\boldsymbol{\theta}|\,\boldsymbol{x})$, which describes the probability distribution of obtaining our model parameters, $\boldsymbol{\theta}$, given an observation, $\boldsymbol{x}$. This characterises the best set of model parameters for describing the given data. This posterior is computed following Bayes' theorem,
\begin{eqnarray}
p(\boldsymbol{\theta}|\,\boldsymbol{x}) = \frac{p(\boldsymbol{x}|\,\boldsymbol{\theta})}{p(\boldsymbol{x})}p(\boldsymbol{\theta})
\end{eqnarray}
where $p(\boldsymbol{x}|\,\boldsymbol{\theta})$ is the likelihood to obtain our observation given our set of model parameters, $p(\boldsymbol{\theta})$ characterises our prior knowledge of reasonable values for our model parameters and $p(\boldsymbol{x})$ is the evidence of the data. 

The basic idea of SBI is to replace the explicit likelihood evaluation with a stochastic simulator of the signal. With this, we generate a training set of data-parameter pairs, $\left[(\boldsymbol{x}_{1},\boldsymbol{\theta}_{1}),...(\boldsymbol{x}_{N},\boldsymbol{\theta}_{N})\right]$ which are drawn from our prior distribution and connects our model parameters to the observed data. We then train a neural network on this data to estimate either the posterior, the likelihood or the likelihood-to-evidence ratio. The advantage of these approaches is that we no longer require any assumptions on the form of the likelihood enabling any complex summary statistic to be explored provided we can compute it in our forward-modelled simulations.

In this work, we perform SBI using \textsc{\small Swyft} \citep{Miller:2022}. Specifically, it performs marginal neural ratio estimation (MNRE; e.g. \citealt{Durkan:2020,Hermans:2021}) to approximate the marginal likelihood-to-evidence ratio for any individual parameter or 2D parameter pair (denoted $\boldsymbol{\tilde\theta}$ to signify any parameter pair; i.e $(\theta_{i}, \theta_{j})$ rather than the likelihood-to-evidence ratio of the entire parameter set. Denoting $r(\boldsymbol{x},\boldsymbol{\tilde\theta})$ to be this marginal likelihood-to-evidence ratio:
\begin{eqnarray}
r(\boldsymbol{x},\boldsymbol{\tilde\theta}) \equiv \frac{p(\boldsymbol{x}|\,\boldsymbol{\tilde\theta})}{p(\boldsymbol{x})} = \frac{p(\boldsymbol{\tilde\theta}|\,\boldsymbol{x})}{p(\boldsymbol{\tilde\theta})} = \frac{p(\boldsymbol{x},\boldsymbol{\tilde\theta})}{p(\boldsymbol{x})p(\boldsymbol{\tilde\theta})},
\end{eqnarray} 
which is the ratio of the probability density for a jointly drawn sample-parameter pair, $\boldsymbol{x},\boldsymbol{\tilde\theta} \sim p(\boldsymbol{x},\boldsymbol{\tilde\theta})$ and a marginally pair $\boldsymbol{x},\boldsymbol{\tilde\theta} \sim p(\boldsymbol{x})p(\boldsymbol{\tilde\theta})$. This ratio is estimated by training a binary classification network, $d_{\phi}(\boldsymbol{x},\boldsymbol{\tilde\theta})$, where $\phi$ describes the network parameters, which distinguishes between two hypotheses: whether the sample-parameter pairs are jointly ($C=1$) or marginally ($C=0$) drawn. The binary classifier is trained using a binary-cross entropy loss function:
\begin{eqnarray}
L[d_{\phi}(\boldsymbol{x},\boldsymbol{\tilde\theta})] &=& -\int {\rm d}\boldsymbol{x}{\rm d}\boldsymbol{\tilde\theta}\left\{ p(\boldsymbol{x},\boldsymbol{\tilde\theta}) {\rm log}d_{\phi}(\boldsymbol{x},\boldsymbol{\tilde\theta}) \right. + \nonumber \\ 
& & \left. p(\boldsymbol{x})p(\boldsymbol{\tilde\theta}){\rm log}\left[ 1- d_{\phi}(\boldsymbol{x},\boldsymbol{\tilde\theta}) \right]\right\},
\end{eqnarray}
which is minimised when $d_{\phi}(\boldsymbol{x},\boldsymbol{\tilde\theta})$ approximates the probability density of the jointly drawn sample-parameter pair (e.g. $C=1$). This returns, 
\begin{eqnarray}
d_{\phi}(\boldsymbol{x},\boldsymbol{\tilde\theta}) = p(C=1|\boldsymbol{x},\boldsymbol{\tilde\theta}) = \frac{p(\boldsymbol{x},\boldsymbol{\tilde\theta})}{p(\boldsymbol{x},\boldsymbol{\tilde\theta}) + p(\boldsymbol{x})p(\boldsymbol{\tilde\theta})} \equiv \sigma[{\rm log}\, r(\boldsymbol{x},\boldsymbol{\tilde\theta})], && \nonumber \\
&&
\end{eqnarray}
where the last equality connects the likelihood-to-evidence ratio, $r$, to the binary classifier, $d_{\phi}$ using the sigmoid function, $\sigma(y) = [1+{\rm e}^{-y}]^{-1}$.

As this approach only learns the marginal likelihood-to-evidence ratio for any parameter pair, $\boldsymbol{\tilde\theta}$, for an $M$ dimensional model we are required to train $M$ 1D and $M(M-1)/2$ 2D networks to fully describe the marginal posterior distribution given an observation. This is because the simulated training set generated by the stochastic simulator inherently contains the variance due to all sampled model parameters. Therefore, the marginalisation over the remaining (nuisance) model parameters is always implicitly performed and thus the training of the binary classifier is limited to at most two dimensions.

\subsection{Simulated data} \label{sec:sim_data}

Our only requirement for SBI is that our stochastic simulator models the complexities of the cosmic 21-cm signal including the observational characteristics of realistic data and that our dataset contains sufficient samples. Below, we summarise the main steps adopted for pipeline based on our previous work \citep[e.g.][Greig et al., in prep]{Greig:2022,Greig:2023}.

We generate 3D realisations of the cosmic 21-cm signal using \cmfst{}, simulated within 250$^{3}$ Mpc$^{3}$ comoving volumes on a 150$^{3}$ grid. The final evolved density fields are downsampled from an initially higher resolution grid, 450$^{3}$, after applying  second-order Lagrange perturbation theory \citep[e.g][]{Scoccimarro:1998p7939}. We track the evolution of the 21-cm signal from $z=25$ down to $z=5.2$ and stitch together the comoving simulation cubes via linear interpolation to generate a 21-cm light-cone. Using the same training set as constructed in Greig et al., (in prep.) we have 150,000 independent realisations of the cosmic 21-cm signal for our forward modelled training set. Note, for this work, non-linear redshift-space distortions (RSDs) were not included in the simulated 21-cm signal \citep[e.g.][]{Mao:2012p7838,Jensen:2013p1389}. Primarily, these RSDs serve to elongate the 21-cm power along the line-of-sight amplifying the anisotropy of the 21-cm signal. As a result, differences between the 2D PS and 1D PS are likely to be underestimated in this work.

Radio interferometers are only sensitive to the spatial fluctuations in the signal and thus the observed data is zero mean distributed. To mimic this, we first split our 21-cm light-cones into equal comoving distance (250 Mpc) chunks. This choice is adopted in-order to measure our 1D and 2D PS using a 3D cubic volume for computational ease. For each of these chunks we then remove the mean signal before adding in the instrumental effects as outlined below.

\subsubsection{Instrumental noise}

To add interferometric noise along with the finite resolution of the instrument to our simulated 21-cm data we use a modified version of the publicly available \textsc{\small Python} module \textsc{\small 21cmSense}\footnote{https://github.com/jpober/21cmSense} \citep{Pober:2013p41,Pober:2014p35}. Provided any antenna configuration \textsc{\small 21cmSense} first generates the corresponding $uv$-visibility tracks for each sampled baseline before gridding for computational efficiency. Specifically for this work, we use the SKA configuration System Baseline Design document\footnote{http://astronomers.skatelescope.org/wp-content/uploads/2016/09/SKA-TEL-SKO-0000422\textunderscore 02\textunderscore SKA1\textunderscore LowConfigurationCoordinates-1.pdf} which includes 512 37.5m antennae stations distributed within a 500m core radius. These stations are modelled assuming a system temperature, $T_{\rm sys} = 1.1T_{\rm sky} + 40~{\rm K}$ and a corresponding sky temperature of $T_{\rm sky} = 60\left(\frac{\nu}{300~{\rm MHz}}\right)^{-2.55}~{\rm K}$ \citep{Thompson2007}. For our setup, we assume a total observing time of 1000 hours based on a single six-hour phase-tracked scan of the sky per night.

Taking the gridded $uv$-visibilities as input, \textsc{\small 21cmSense} then computes the total thermal noise power, $P_{\rm N}(k)$;
\begin{eqnarray} \label{eq:NoisePS}
P_{\rm N}(k) \approx X^{2}Y\frac{\Omega^{\prime}}{2t}T^{2}_{\rm sys},
\end{eqnarray} 
where $X^{2}Y$ performs the cosmological conversions between observing bandwidth, frequency and co-moving distance, $\Omega^{\prime}$ is a beam-dependent factor derived by \citet{Parsons:2014p781} and $t$ is the total observing time.

As we are interested in 3D noise realisations rather than the 1D total noise power we perform the following modifications:
\begin{itemize}
\item We first 3D Fourier transform the input (simulated) mean removed 21-cm data cube
\item We then filter this cube using the gridded $uv$-visibilities for the SKA computed by \textsc{\small 21cmSense}. Cells with finite $uv$-coverage are multiplied by unity, all others are set to zero
\item At each cell we then determine the amplitude of the thermal noise, $P_{\rm N}(k_{x}, k_{y}, k_{z})$, using Equation~\ref{eq:NoisePS} where $k_x$ and $k_y$ correspond to the two transverse (on sky) directions and $k_z$ is the line-of-sight direction
\item We then add random noise (zero mean with variance based on the power spectrum amplitude in the cell) to each cell to mimic the effect of thermal noise
\item Finally, we then 3D inverse Fourier transform back to obtain our noisy 21-cm data.
\end{itemize}

\subsubsection{The foreground wedge} \label{sec:wedge}

Unfortunately, individual $uv$ visibilities from a radio interferometer baseline are frequency dependent. This means that the line-of-sight (frequency dependent) power can leak into the transverse (frequency independent) Fourier modes resulting in a well-defined contaminated `wedge' in cylindrical 2D Fourier space \citep{Datta:2010p2792,Vedantham:2012p2801,Morales:2012p2828,Parsons:2012p2833,Trott:2012p2834,Thyagarajan:2013p2851,Liu:2014p3465,Liu:2014p3466,Thyagarajan:2015p7294,Thyagarajan:2015p7298,Pober:2016p7301,Murray:2018}. This gives rise to two separate philosophies for dealing with this wedge contamination: foreground removal and foreground avoidance. 

In the first case, we assume that we can mitigate or `clean' these contaminated modes (see e.g.\ \citealt{Chapman:2019} for a review, or by using machine learning \citealt{Gagnon-Hartman:2021}) enabling us to recover and use the entire 21-cm signal. In the latter case, we conservatively avoid this wedge contaminated region of Fourier space and only use the `clean' Fourier modes located above this `wedge'. In this work, we shall consider both scenarios when exploring the 2D PS for parameter inference.

While the foreground removal case utilises the full simulated 21-cm data, performing wedge avoidance requires an additional step to those discussed in the previous section. The boundary defining this foreground `wedge' in 2D Fourier space is given by,
\begin{eqnarray} \label{eq:wedge}
k_{\parallel} =  mk_{\perp} + b
\end{eqnarray} 
where $k_{\parallel}$ and $k_{\perp}$ are the line-of-sight and transverse Fourier modes, $b$ is a additive buffer which we assume to be $\Delta k_{\parallel} = 0.1 \,h$~Mpc$^{-1}$ which accounts for bleeding of noise extending beyond the horizon limit and $m$ is the gradient of this boundary given by
\begin{eqnarray}
m = \frac{D_{\rm C}H_{0}E(z){\rm sin}(\theta)}{c(1+z)}.
\end{eqnarray} 
This boundary depends on the comoving distance, $D_{\rm C}$, the Hubble constant, $H_{0}$, cosmological factor $E(z) = \sqrt{\Omega_{\rm m}(1+z)^{3} + \Omega_{\Lambda}}$ and ${\rm sin}(\theta)$ denotes the observed viewing angle for our observation, for which we assume as $\theta = \pi/2$ (i.e. a zenith pointing observation).
 
In order to account for the foreground wedge, we must remove the Fourier modes from below the wedge. Therefore, after 3D Fourier transforming our input 3D 21-cm data cube, we first zero all modes that fall below this foreground `wedge' before adding the thermal noise for all modes above the wedge.

\subsection{Mock 21-cm Observation}

\begin{figure*}
	\includegraphics[trim = 0cm 0.6cm 0cm 0.5cm, scale = 0.44]{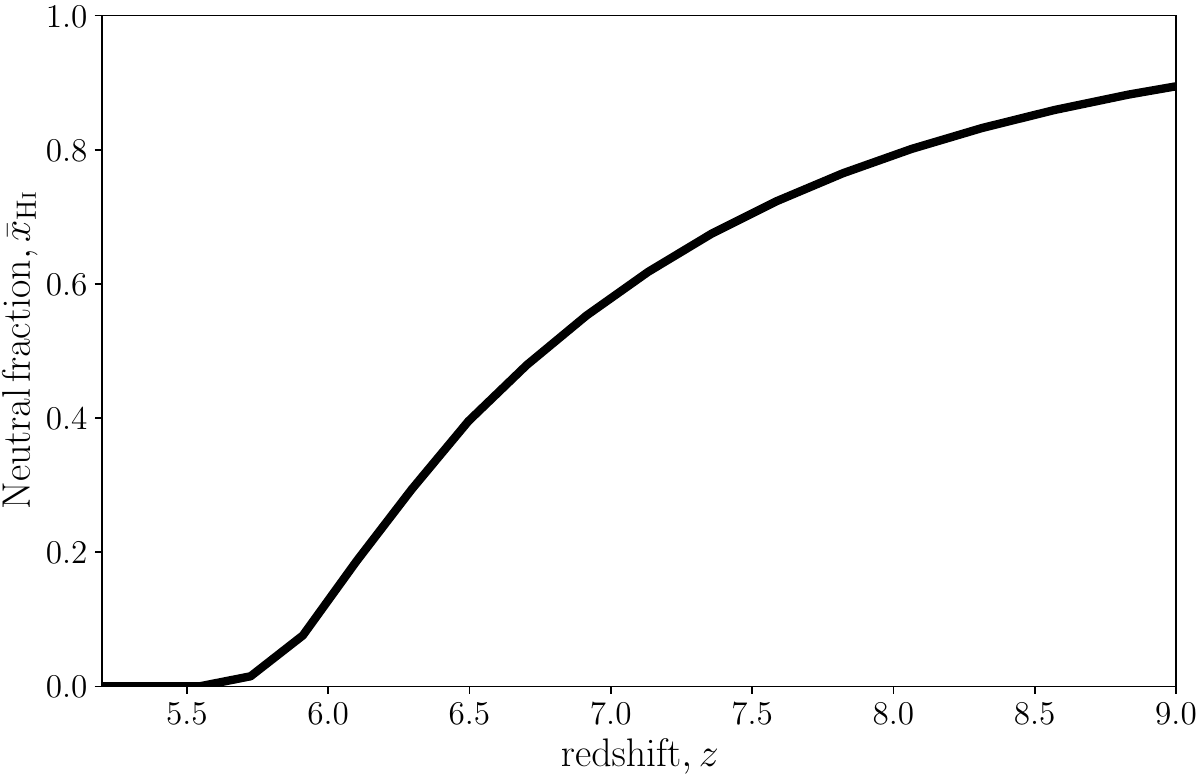}
	\includegraphics[trim = 0.3cm 0.6cm 0cm 0.5cm, scale = 0.44]{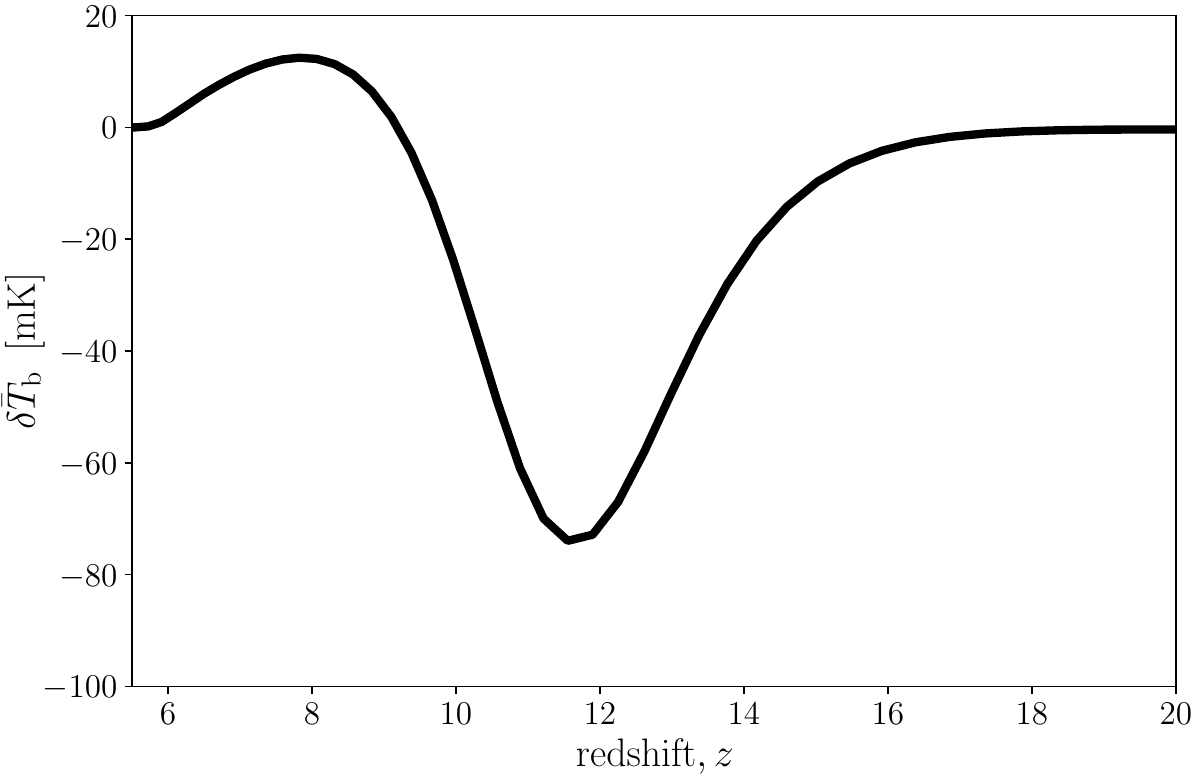}	
    \caption{The volume averaged IGM neutral fraction (left panel) and the mean brightness temperature signal (right panel) for our fiducial astrophysical parameter set used for constructing our mock observation.}
    \label{fig:history}
\end{figure*}

Exploring the 2D PS for parameter inference requires the construction of a mock observation of the 21-cm signal. For this, we assume the same fiducial parameter set as in Greig et al. (in prep). Namely, we select parameters for our UV galaxies in line with the recovered model of \citet{Qin:2021} based on \lya{} forest observations by \citet{Bosman:2018}. Below we summarise the individual model parameter values along with their associated flat prior ranges and in Figure~\ref{fig:history} we provide the volume averaged IGM neutral fraction and mean brightness temperature as a function of redshift:
\begin{itemize}
\item ${\rm log}_{10}(f_{\ast,10}) = -1.10; \in [-3.0,0.0]$
\item $\alpha_{\ast} = 0.5; \in [-0.5,1.0]$
\item ${\rm log}_{10}(f_{\rm esc,10}) = -1.30; \in [-3.0,0.0]$
\item $\alpha_{\ast} = -0.35; \in [-1.0,0.5]$
\item ${\rm log}_{10}(M_{\rm turn}) = 8.55; \in [8.0,10.0]$
\item $t_{\ast} = 0.5; \in [0.05,1.0]$
\item $L_{{\rm X}<2\,{\rm keV}}/{\rm SFR} = 40.50; \in [38.0,42.0]$
\item $E_{0} = 0.5; \in [0.1,1.5]$
\end{itemize}

\begin{figure*}
	\includegraphics[trim = -0.05cm 0.2cm 0cm 0cm, scale = 0.88]{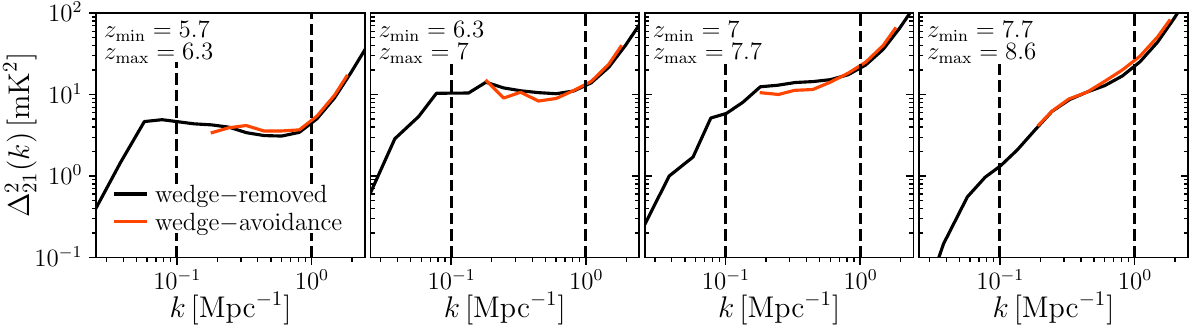}
	\includegraphics[trim = 0.2cm 0.2cm 0cm 0cm, scale = 0.9]{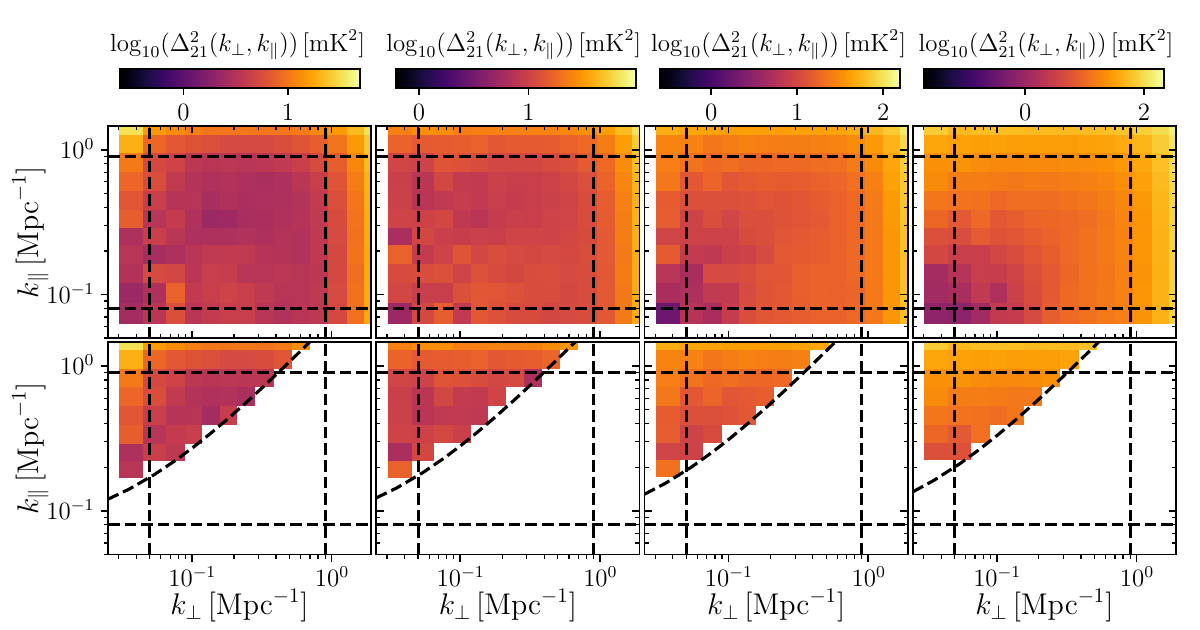}
    \caption{Comparison of the 1D PS and the 2D PS from four different redshift ranges obtained from our simulated 21-cm light-cone. \textit{Top row:} The 1D PS assuming perfect foreground removal (black curve) and after wedge avoidance (red curve). The vertical dashed lines at $k=0.1$ and $1.0$~Mpc$^{-1}$ correspond to the region within which we use for parameter inference. \textit{Middle row:} the 2D PS assuming perfect foreground removal. The vertical dashed lines correspond to $k_{\perp}=0.05$ and $0.9$~Mpc$^{-1}$ and the horizontal dashed lines correspond to $k_{\parallel}=0.08$ and $0.9$~Mpc$^{-1}$. We use all 2D PS information bounded within these regions for our parameter inference. \textit{Bottom row:} the 2D PS after performing wedge avoidance (observing only modes above the wedge denoted by the diagonal black dashed line).}
    \label{fig:mock}
\end{figure*}

In Figure~\ref{fig:mock} we compare the 1D and 2D PS for the first four redshift ranges extracted from our mock observation of the 21-cm light-cone. Additionally, we demonstrate the differences in measured PS as a result of the two distinct treatments of the astrophysical foregrounds. For the 1D PS demonstrated in the top row, we distinguish between perfect foreground wedge removal (black) and wedge-avoidance (red), respectively. The black vertical dashed lines correspond to the region of the 1D PS between which we use for performing astrophysical parameter inference, namely $k=0.1$ and $1.0$~Mpc$^{-1}$. The impact of ignoring the foreground contaminated wedge region is clearly evident here, restricting the range of Fourier modes accessible for performing our inference. Beyond the visible removal of modes, there will also be less spherically averaged modes per $k$-bin for the wedge avoidance case, which will also lead to an increase in the corresponding sample variance uncertainty resulting in broadened inferred astrophysical posteriors.

In the middle panel of Figure~\ref{fig:mock} we provide the 2D PS assuming perfect foreground removal, whereas the bottom panel corresponds to the wedge-avoidance scenario. The vertical and horizontal dashed lines correspond to the boundaries for the Fourier modes we shall consider for our parameter inference from the 2D PS. These are selected to roughly balance modes that are adequately sampled by our simulation volume and also to correspond to the same scales as used in our inference pipeline for the 1D PS. Namely, we consider $k_{\perp}=0.05$ and $0.9$~Mpc$^{-1}$ and $k_{\parallel}=0.08$ and $0.9$~Mpc$^{-1}$. Note, although with these boundaries it is possible to sample modes at $k>1$~Mpc$^{-1}$ (where $k = \sqrt{(k^{2}_{\perp}+k^{2}_{\parallel})}$) these modes are the most severely affected by instrumental thermal noise and thus will not provide much additional constraining power.

From these two panels it is immediately clear the significant impact that the foreground wedge has on our ability to measure the 21-cm signal. In terms of the 2D PS, the foreground wedge contaminates well over 60 per cent of the Fourier modes contained within our boundaries to be used for parameter inference. Further, this only gets worse for increasing redshifts as the wedge boundary is redshift dependent. However, what matters is where the information is lost. The vast majority of the information is lost for large $k_{\perp}$, which is less sensitive to the astrophysical parameters. In the case of the 1D PS, most of the constraining power comes from the `knee' like feature around $k\sim0.1$~Mpc$^{-1}$ \citep[see][]{Greig:2015p3675} which corresponds to the typical sizes of the ionised regions. Since the 2D PS still has reasonable sampling of these modes, that is for low $k_{\perp}$, we should not see such a drastic reduction in constraining power between the two foreground mitigation scenarios as we are still sensitive to the morphological information both during the EoR and in the Epoch of X-ray Heating (EoH). It will be instructive to quantify the relative difference between the two mitigation strategies.

Note there are several competing effects leading to our inability to access information below $k\sim0.15$~Mpc$^{-1}$ for the 1D PS. Simply increasing the simulation size does not immediately alleviate the issue. Firstly, as we aim to mimic realistic observations of the 21-cm signal from a radio interferometer, when computing the 1D PS we do not consider the case when $k_{\perp} = 0$ (i.e. $k = k_{\parallel}$). These modes are not visible to radio interferometers as the minimum available baseline is the diameter of the receiving element (dish or antennae station). This, coupled with the definition of the foreground wedge severely limits the spherically binned Fourier modes below $k\sim0.15$~Mpc$^{-1}$. For instance, we have our horizon buffer at $0.1 h$~Mpc$^{-1}$, which sets the minimum allowed for $k_{\parallel}$ (Equation~\ref{eq:wedge}). Therefore, to obtain $k\sim0.1$~Mpc$^{-1}$ for wedge-avoidance, we require $k_{\perp}\lesssim0.015$~Mpc$^{-1}$ corresponding to a simulation with side-length of at least $\sim$420~Mpc. However, this is for one single mode, to have a reasonable statistical sampling of $k_{\perp}\lesssim0.015$~Mpc$^{-1}$ we would require at least $2-3$ times larger side-lengths, corresponding to $\gtrsim1$~Gpc. Generating such large simulation volumes for parameter inference are infeasible. Of course, this would be less severe if we removed this additive horizon buffer. Importantly, since we can obtain a reasonable statistical sampling of Fourier modes for our 2D PS in the case of perfect foreground removal, our simulation volumes are sufficient for this analysis.

\section{Parameter inference with the 2D PS} \label{sec:results}

\subsection{Estimating the posteriors with SBI}

In \textsc{Swyft}, to obtain our desired marginal posterior distributions for our astrophysical parameters we need to construct neural networks to learn the likelihood-to-evidence ratios given our mock observation of the 21-cm signal. Following our previous work, Greig et al. (in prep), we split our simulated 3D 21-cm light-cone into ten equal co-moving chunks spanning from $z=5.7$ to $z=18.1$, from which we measure either the 1D or 2D PS. Although the SKA is designed to be sensitive down to 50 MHz ($z\sim27.8$), at these redshifts the thermal noise dominates over our fiducial model, therefore, we limit the redshift dimension for our data to $z\leq18$. Restricting our inference to Fourier modes between $k=0.1$~Mpc$^{-1}$ and $k=1.0$~Mpc$^{-1}$, resulted in a total of 60 1D PS data-points when considering wedge avoidance. These were simply taken as the input layer to a three-layered fully connected neural network consisting of 256 neurons. That is, we do not use an embedding network to reduce the dimensionality of the input data prior to the fully connected neural network. For the 1D PS under perfect foreground removal, we instead recover eight Fourier modes within our defined boundary, resulting in a total of 80 1D PS data-points. For this, we use the same network architecture, with these 80 data-points being the input layer.

For the 2D PS, as one would expect we have considerably more input data. Considering wedge-avoidance, and our corresponding Fourier cuts, we obtain 161 data-points for the 2D PS. This is not overly restrictive computationally and thus we retain this data as a linear input layer for our three layered neural network. On the other hand, assuming perfect foreground removal, we have a total of 900 datapoints. In this work, we choose to apply a linear transformation to this data to reduce it down to 256 features with which we take as our 1D input to our three layered neural network, with 512 neurons per layer. That is, we adopt this linear transformation as an embedding network. However, given that the 2D PS naturally lends itself to a 2D image representation of the data, one could instead apply a convolutional neural network (CNN) as our embedding network to more optimally extract the features within the data. For example, \citet{Breitman:2024} found that applying a CNN to the 1D PS data represented as a 2D image ($k,z$) improved the overall performance of their emulator, implying better relative performance at feature extraction. Nevertheless, after exploring several different network architectures, we found this linearisation of the data to be sufficient to extract the data, given the number of Fourier modes that are dominated by thermal noise. However, in future we will return to this to perform a more rigorous exploration of optimal network architectures for extracting the relevant features of our data.

\subsection{Perfect foreground removal}

\begin{figure*}
	\includegraphics[trim = 0.2cm 0.2cm 0cm 0.5cm, scale = 0.92]{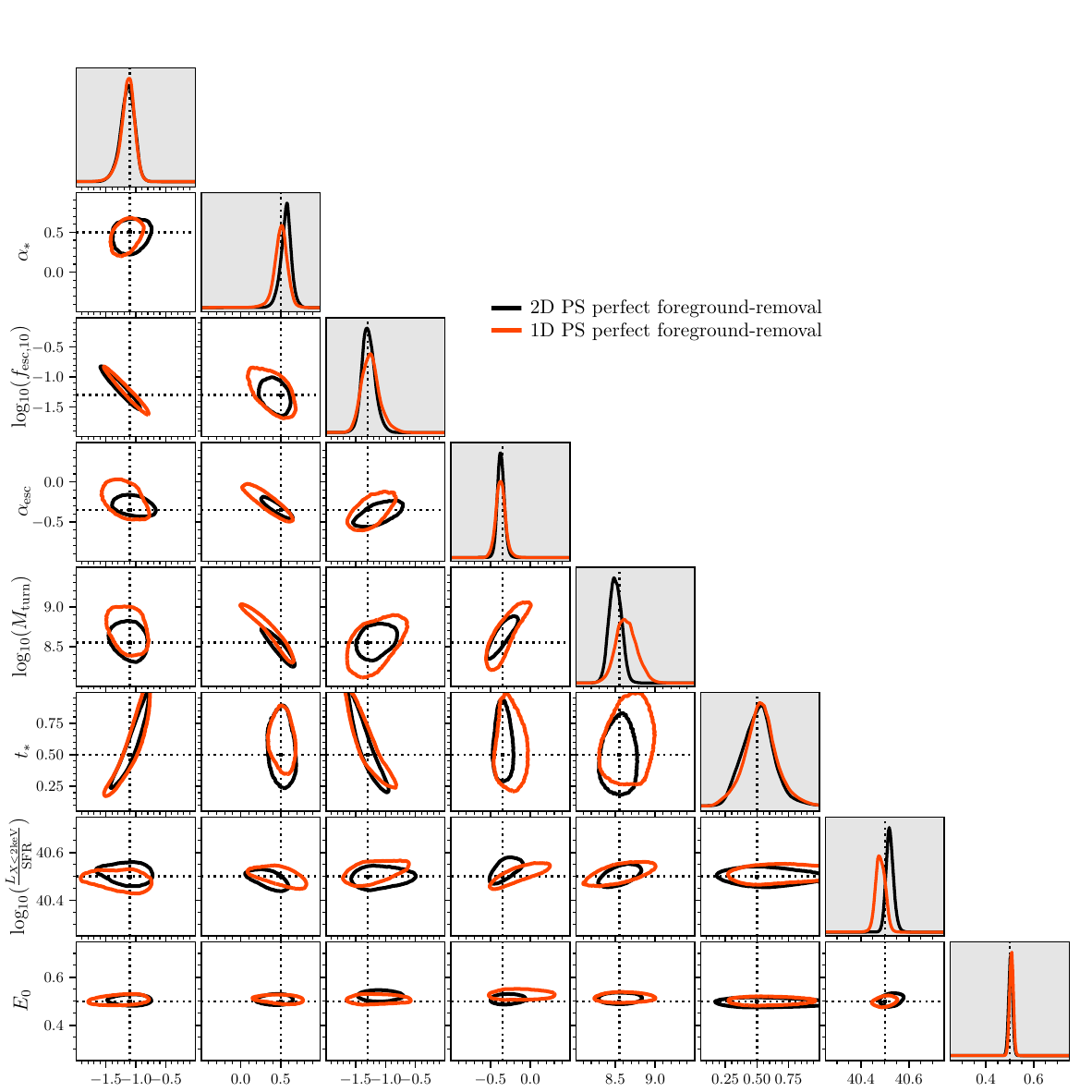}
    \caption{The recovered one and two dimensional marginalised posteriors on our astrophysical parameters assuming a mock 1000 hr observation of the 21-cm signal assuming perfect foreground removal with the SKA. Black curves correspond to using the 2D cylindrically averaged PS (2D PS) whereas the red curves correspond to the 3D spherically averaged PS (1D PS). The 2D contours below the diagonal correspond to the 95th percentiles. The vertical and horizontal black dashed lines denote our fiducial astrophysical parameter set.}
    \label{fig:wr}
\end{figure*}

First, we consider the somewhat optimistic case of perfectly removing foreground contamination enabling the full use of the Fourier information (see e.g. Figure~\ref{fig:mock}). In Figure~\ref{fig:wr}, we present the 1D and 2D marginalised posteriors following our SBI approach with \textsc{Swyft} for our mock 21-cm observation. For this we demonstrate the resultant posteriors for the 2D (1D) PS by the black (red) curves, respectively. Below the marginalised 1D PDFs along the diagonal we demonstrate the 95th percentile joint 2D posteriors. In Table~\ref{tab:results} we summarise the recovered constraints and marginalised 68th percentile uncertainties for our eight astrophysical model parameters. Additionally, in Appendix~\ref{sec:coverage} we demonstrate our trained network coverage demonstrating its convergence.

\begin{table*}
 \caption{A summary of the recovered astrophysical parameter constraints plus 68th marginalised uncertainties obtained following SBI on our mock observation of the 21-cm signal using either the 1D or 2D PS. We also consider two foreground mitigation strategies: (i) perfect foreground removal and (ii) foreground avoidance for a 1000 hr observation using the SKA. Finally, we consider the improvements in the constraining power following the inclusion of UV LFs (see text for further details).}
 \label{tab:results}
\begin{tabular}{ccccccccc}
  \hline
 & ${\rm log}_{10}(f_{\ast,10})$ & $\alpha_{\ast}$ & ${\rm log}_{10}(f_{{\rm esc},10})$ & $\alpha_{\rm esc}$ & ${\rm log}_{10}(M_{\rm turn})$ & $t_{\ast}$ & ${\rm log}_{10}\frac{L_{X<2\,{\rm keV}}}{\rm SFR}$ & $E_{0}$\\
 & & & & & ($M_{\odot}$) & & (erg s$^{-1}$ $M^{-1}_{\odot}$ yr$^{-1}$) & (keV) \\
  \hline
Mock Observation & -1.1 & 0.5 & -1.30 & -0.35 & 8.55 & 0.5 & 40.5 & 0.5 \\  
  \hline
foreground removal &  & \\  
  \hline
1D PS & -1.12$\substack{+0.10\\-0.11}$  & 0.51$\substack{+0.07\\-0.08}$ & -1.26$\substack{+0.12\\-0.16}$  & -0.37$\substack{+0.06\\-0.06}$  & 8.61$\substack{+0.14\\-0.11}$  & 0.53$\substack{+0.12\\-0.12}$  & 40.48$\substack{+0.02\\-0.02}$  & 0.51$\substack{+0.01\\-0.01}$  \\
2D PS & -1.13$\substack{+0.11\\-0.11}$  & 0.55$\substack{+0.06\\-0.07}$ & -1.31$\substack{+0.11\\-0.09}$  & -0.37$\substack{+0.04\\-0.04}$  & 8.49$\substack{+0.08\\-0.07}$  & 0.54$\substack{+0.09\\-0.14}$  & 40.51$\substack{+0.02\\-0.01}$  & 0.50$\substack{+0.01\\-0.01}$ \\
\hline
1D PS + UV LFs & -1.13$\substack{+0.10\\-0.11}$  & 0.49$\substack{+0.05\\-0.06}$ & -1.29$\substack{+0.08\\-0.10}$  & -0.31$\substack{+0.04\\-0.04}$  & 8.57$\substack{+0.07\\-0.07}$  & 0.60$\substack{+0.13\\-0.09}$  & 40.50$\substack{+0.01\\-0.01}$  & 0.51$\substack{+0.01\\-0.01}$  \\
2D PS + UV LFs  & -1.09$\substack{+0.10\\-0.10}$  & 0.46$\substack{+0.06\\-0.05}$ & -1.32$\substack{+0.10\\-0.12}$  & -0.34$\substack{+0.03\\-0.04}$  & 8.60$\substack{+0.07\\-0.07}$  & 0.46$\substack{+0.13\\-0.13}$  & 40.50$\substack{+0.01\\-0.01}$  & 0.51$\substack{+0.01\\-0.01}$  \\
  \hline
foreground avoidance &  & \\  
  \hline
1D PS & -1.14$\substack{+0.10\\-0.13}$  & 0.48$\substack{+0.13\\-0.19}$ & -1.25$\substack{+0.12\\-0.14}$  & -0.29$\substack{+0.14\\-0.08}$  & 8.60$\substack{+0.27\\-0.15}$  & 0.48$\substack{+0.14\\-0.15}$  & 40.47$\substack{+0.03\\-0.03}$  & 0.49$\substack{+0.03\\-0.02}$  \\
2D PS & -1.10$\substack{+0.11\\-0.12}$  & 0.51$\substack{+0.08\\-0.11}$ & -1.24$\substack{+0.11\\-0.15}$  & -0.27$\substack{+0.10\\-0.11}$  & 8.43$\substack{+0.18\\-0.15}$  & 0.53$\substack{+0.16\\-0.10}$  & 40.48$\substack{+0.03\\-0.02}$  & 0.48$\substack{+0.04\\-0.03}$  \\  
\hline
1D PS + UV LFs & -1.11$\substack{+0.13\\-0.13}$  & 0.45$\substack{+0.08\\-0.09}$ & -1.20$\substack{+0.13\\-0.11}$  & -0.30$\substack{+0.05\\-0.05}$  & 8.47$\substack{+0.16\\-0.12}$  & 0.58$\substack{+0.16\\-0.12}$  & 40.49$\substack{+0.03\\-0.02}$  & 0.49$\substack{+0.04\\-0.03}$  \\
2D PS + UV LFs  & -1.13$\substack{+0.11\\-0.13}$  & 0.36$\substack{+0.09\\-0.06}$ & -1.30$\substack{+0.12\\-0.11}$  & -0.30$\substack{+0.05\\-0.05}$  & 8.49$\substack{+0.15\\-0.13}$  & 0.57$\substack{+0.15\\-0.11}$  & 40.51$\substack{+0.03\\-0.03}$  & 0.45$\substack{+0.03\\-0.03}$  \\
  \hline
 \end{tabular}
\end{table*}

The 2D PS outperforms the 1D PS as evident by the narrower marginalised posteriors between the two summary statistics. However, the relative improvements are relatively modest. Based on the 68th percentile marginalised uncertainties we see on average improvements of approximately (5, 15, 30, 30, 40, 20) per cent for ($f_{\ast,10}$, $\alpha_{\ast}$, $f_{{\rm esc},10}$, $\alpha_{\rm esc}$, $M_{\rm turn}$, $L_{X<2\,{\rm keV}}/{\rm SFR}$) with no improvement for $t_{\ast}$ or $E_{0}$. Recall, in this work we do not include non-linear RDSs in our simulations of the 21-cm signal, thus these differences likely underestimate the actual differences when RSDs are included which serve to amplify the anisotropy of the 21-cm signal along the line-of-sight, $k_{\parallel}$. Nevertheless, these modest improvements are consistent with the Fisher Matrix expectations of \citet{Prelogovic:2024} who predict relative improvements of 15 per cent on the variance of the individual parameters based on the factor of $\sim2$ improvement in the total Fisher information. Note, we find little to no improvement in the X-ray parameters between the 2D PS and 1D PS. Likely, this is due to the selection of only two free X-ray parameters, $L_{X<2\,{\rm keV}}/{\rm SFR}$ and $E_{0}$ in our model. $L_{X<2\,{\rm keV}}/{\rm SFR}$ is constrained by the PS amplitude and is relatively independent of the EoH morphology. By only having one morphological X-ray parameter, $E_{0}$, combined with the increasing thermal noise to higher redshifts we limit the ability for the 2D PS to outperform the 1D PS. If we were to additionally consider the spectral index of the X-ray photons, $\alpha_{\rm X}$, as a free parameter, which is degenerate with $E_{0}$ then we would anticipate the 2D PS outperforming the 1D PS for this parameter combination due to the additional 2D spatial information provided by the 2D PS. Although the relative improvement would still depend on the thermal noise amplitude.

These improvements in the constraining power arise due to the distinction of the Fourier information into their transverse ($k_{\perp}$) and redshift evolving ($k_{\parallel}$) components. Although the relative noise in each individual $k_{\perp},k_{\parallel}$ bin increases due to the larger sample variance relative to the spherically averaged $k$-bins of the 1D PS, the anisotropic nature of the 21-cm signal yields additional information (see e.g. Figure~\ref{fig:mock}). By having the transverse spatial information independent of redshift we are more sensitive to the spatial morphology during the EoR and EoH. That is, sampling $k_{\perp}$ for a specific $k_{\parallel}$ provides unique information about the relative amplitudes of the spatial fluctuations as a function of redshift, providing more fine-grained detail than the 1D PS, which averages the anisotropic signal into the `knee'-like feature at $k\sim0.1$~Mpc$^{-1}$.

For example, since we are more sensitive to the spatial morphology (i.e. distribution of the ionised regions as a function of scale and redshift), we recover improved constraints on the parameters that control the typical sizes of the ionised regions. Namely $M_{\rm turn}$ which defines the characteristic masses of the star-forming galaxies and both the normalisations and mass-dependent power law indices of $f_{\ast}$ and $f_{\rm esc}$ which control the production of UV ionising photons and their escape into the IGM to drive ionisations.

\subsection{Foreground avoidance}

\begin{figure*}
	\includegraphics[trim = 0.2cm 0.2cm 0cm 0.5cm, scale = 0.92]{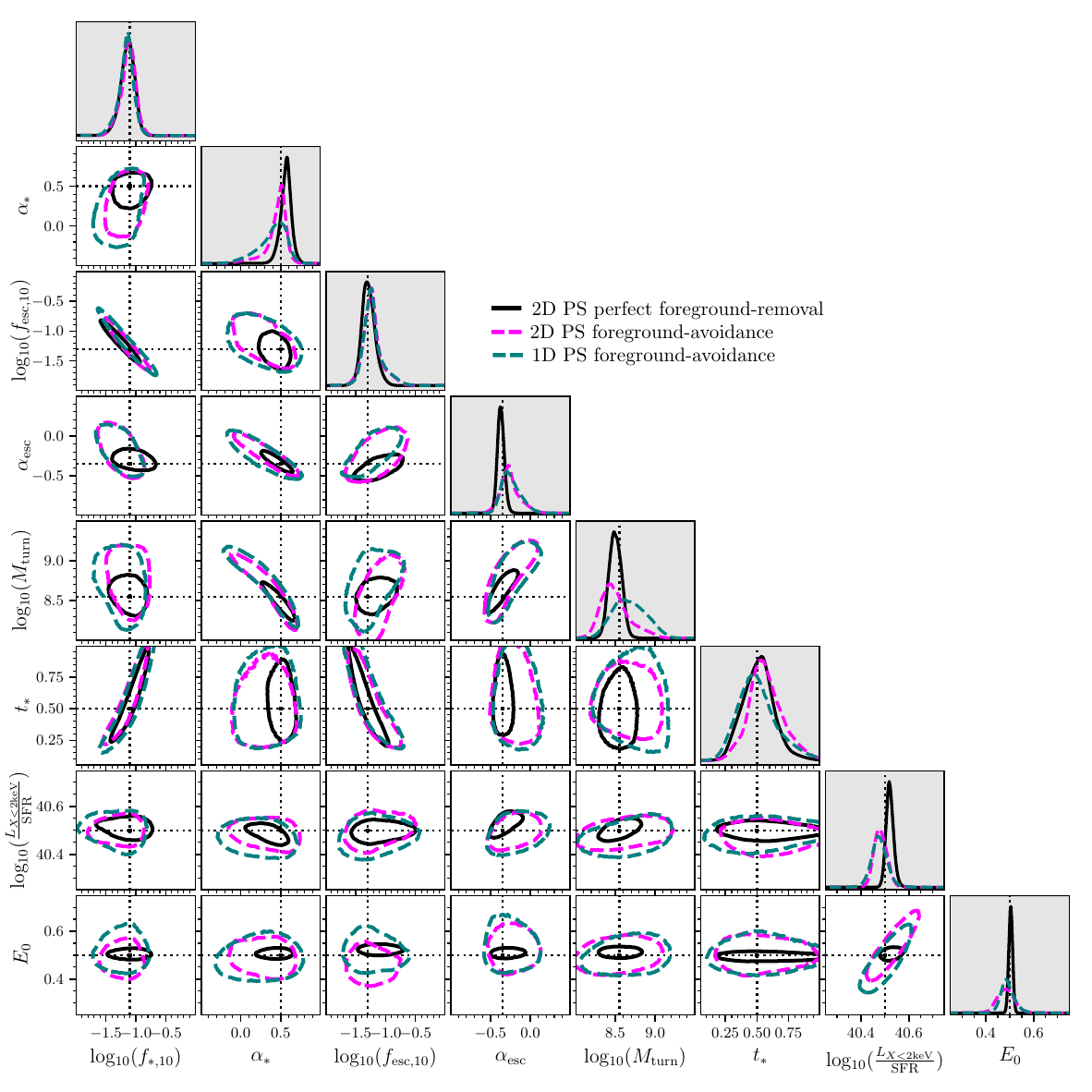}
    \caption{The same as Figure~\ref{fig:wr} except now considering a 1000 hr observation with the SKA assuming foreground wedge avoidance. The magenta (teal) dashed contours correspond to the 2D (1D) PS, whereas the black contours represent the 2D PS assuming foreground removal for comparison.}
    \label{fig:wa}
\end{figure*}

In Figure~\ref{fig:wa}, we now show the marginalised 1D and 2D posteriors for the same mock 21-cm observation of the 1D (teal dashed) and 2D PS (magenta dashed) except when assuming foreground wedge avoidance for a 1000h observation with the SKA. For comparison, we also show the posteriors for the 2D PS assuming perfect foreground removal. Again, we provide the recovered constraints and 68th percentile uncertainties in Table~\ref{tab:results}. 

As one would expect, considering foreground avoidance results in reduced constraining power relative to perfect foreground removal. However, the 2D PS still outperforms the 1D PS, albeit to a slightly lesser extent. Averaging over the marginalised 68th percentile uncertainties, we recover improvements of approximately (40, 10, 25, 15, 15) per cent for ($\alpha_{\ast}$, $\alpha_{\rm esc}$, $M_{\rm turn}$, $t_{\ast}$, $L_{X<2\,{\rm keV}}/{\rm SFR}$), with little to no improvement for the remaining parameters. In short, the 2D PS always outperform the 1D PS irrespective of the foreground removal strategy. Note, the slightly different selection of parameters that recover slight improvements (e.g. $f_{\ast,10}$, $f_{{\rm esc},10}$ and $t_{\ast}$) between the two foreground strategies are due to the strength of the complex parameter degeneracies and the fairly modest actual improvements in these specific parameters. 

After considering foreground avoidance, we do not see an improvement in $f_{\ast,10}$ or $f_{{\rm esc},10}$, with the only improvements in $f_{\ast}$ and $f_{\rm esc}$ coming from their power-law mass dependence (e.g. $\alpha_{\ast}$ and $\alpha_{\rm esc}$). Nevertheless, we still recover improvements of $\sim10-40$ percent for $M_{\rm turn}$ and these power-law indices which highlights that even when applying foreground avoidance, we still pick up additional information from the spatial morphology of the 21-cm signal due to how the 2D PS distinguishes the structural information from that of redshift evolution. This is despite the fact that after applying foreground avoidance we loose more than 60-80 per cent of the 2D PS Fourier modes (see e.g. Figure~\ref{fig:mock}). However, predominately this information loss is for larger $k_{\perp}$'s, with still relatively decent sampling of $k_{\perp}$ at $\lesssim0.1$~Mpc$^{-1}$ where we predominately extract most of the constraining information \citep[][]{Greig:2015p3675}. At these scales, we are still recovering the redshift evolution of the 21-cm signal (e.g. $k_{\parallel})$, therefore we remain sensitive to how the EoR morphology evolves with redshift. This enables the still relatively strong constraints on our astrophysical parameters.

Relative to perfect foreground removal, for the 2D PS we determine increases in the marginalised 68th percentiles of approximately (5, 50, 30, 275, 200, 5, 70, 300) per cent for ($f_{\ast,10}$, $\alpha_{\ast}$, $f_{{\rm esc},10}$, $\alpha_{\rm esc}$, $M_{\rm turn}$, $t_{\ast}$, $L_{X<2\,{\rm keV}}/{\rm SFR}$, $E_{0}$) by considering foreground avoidance. Over our entire mock 21-cm observation with the 2D PS, foreground avoidance results in a loss of $\sim60-80$ per cent of the 2D Fourier modes owing to the redshift dependence of the foreground wedge. Therefore, despite the loss in over $\sim5$ times the amount of information, we do not exhibit such severe losses in constraining power. What is important is not the total amount of information lost, but rather where this information is lost. As highlighted above, we still access the redshift evolution of the 21-cm PS on those scales most sensitive to the EoR (e.g. $k_{\perp}\sim0.1$~Mpc$^{-1}$). As a result, we recover relatively more modest losses in constraining power on our EoR parameters. Nevertheless, the loss in information below the wedge at moderate scales, $k_{\perp}\sim0.5$ does limit our ability to constrain $M_{\rm turn}$ which drives the increased uncertainties in the power-law indices. For the X-ray parameters, the relative losses are more significant, and this is due to the increasing amplitude of the wedge during the heating epoch removing more spatial information (the wedge moves vertically upward in Figure~\ref{fig:mock} for increasing redshift). Thus we have considerably less spatial information during the EoH heating. However, despite these uncertainties increasing by up to a factor of $\sim3$ the X-ray parameters are still very strongly constrained, highlighting how sensitive the X-ray parameters are tied to the amplitude of the 21-cm signal.

Repeating this analysis for the 1D PS, we recover increases in the marginalised 68th percentiles of approximately (10, 200, 10, 200, 70, 25, 60, 300) per cent for ($f_{\ast,10}$, $\alpha_{\ast}$, $f_{{\rm esc},10}$, $\alpha_{\rm esc}$, $M_{\rm turn}$, $t_{\ast}$, $L_{X<2\,{\rm keV}}/{\rm SFR}$, $E_{0}$) by considering foreground avoidance instead of foreground removal. These relative increases are comparable in amplitude to those for the 2D PS, as one would expect. Again, this highlights that it is not the amount of information lost, rather where the information is lost relative to where the 21-cm signal is most sensitive.

\subsection{Mock 2D PS observation with UV LFs}

\begin{figure*}
	\includegraphics[trim = 0.25cm 0.2cm 0cm 0.5cm, scale = 0.9]{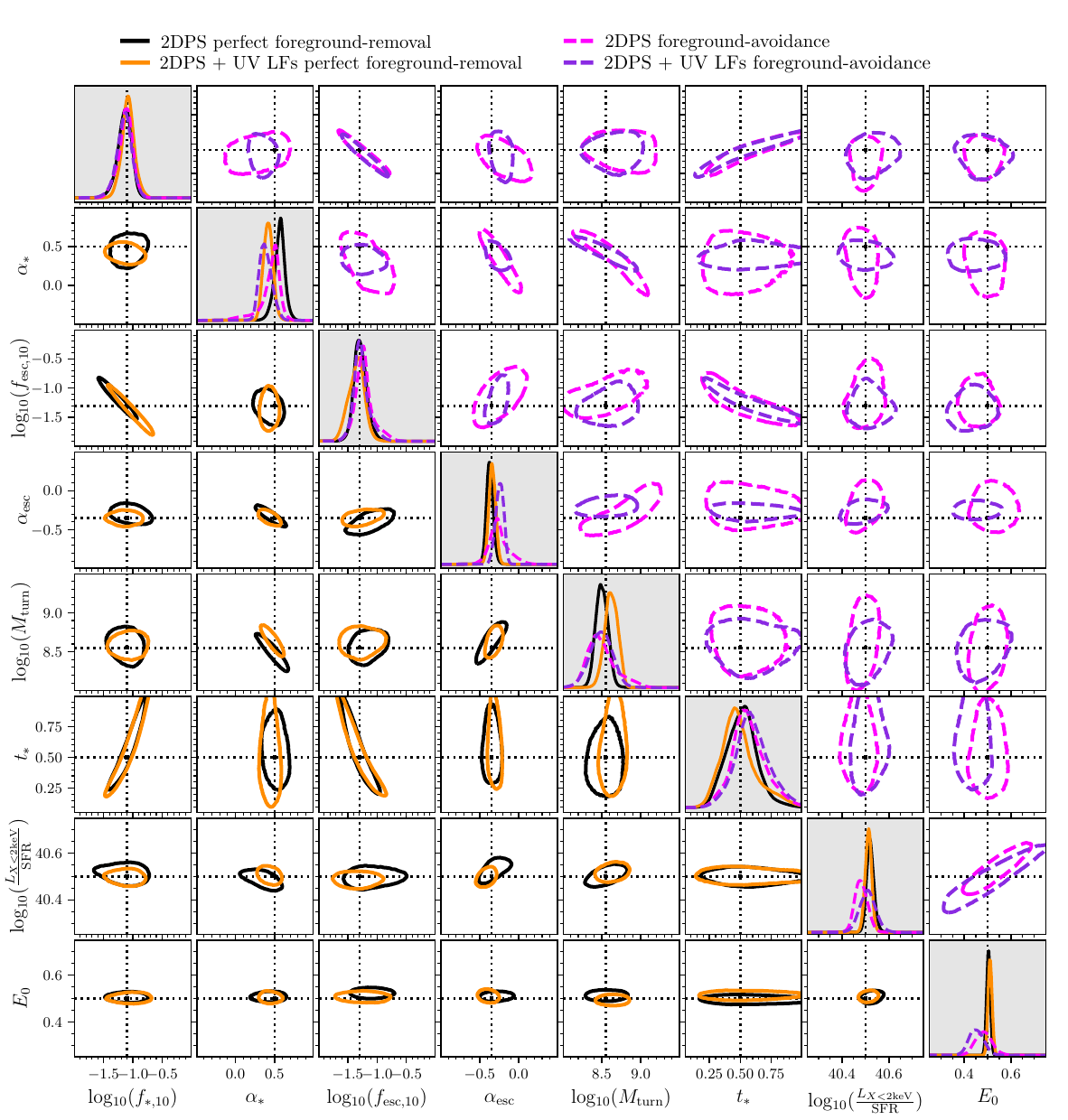}
    \caption{The recovered one and two dimensional marginalised posteriors on our astrophysical parameters assuming a mock 1000 hr observation of the 21-cm signal using the 2D PS from the SKA combined with UV LFs at $z=6-10$. Below the diagonal, we consider the case of perfect foreground removal, with the orange (black) curves corresponding to with (without) the UV LFs, respectively. Above the diagonal, we present the results instead assuming foreground avoidance, with the magenta (purple) dashed curves denoting observations with (without) the UV LFs. In all cases, the 2D contours represent the joint 95th percentile marginalised uncertainties. The vertical and horizontal black dashed lines denote our fiducial astrophysical parameter set.}
    \label{fig:wa-combined}
\end{figure*}

Thus far, we have only considered the relative improvements in our astrophysical parameter constraints when considering the 2D PS compared to the 1D PS. However, one can also include additional constraining information from alternative probes of the reionisation epoch, such as observed UV luminosity functions (LFs). This simply requires concatenating the UV LF data to the existing PS data and passing this information into \textsc{Swyft} and retraining the ratio networks. In the case of the 1D PS, the role of the UV LFs is to break the degeneracy between $f_{\ast}$ and $f_{\rm esc}$ improving the constraining power on these parameters and consequently also on $M_{\rm turn}$ \citep{Park:2019}. Since these same parameters are more strongly constrained with the 2D PS relative to the 1D PS, it will thus be illustrative to consider a joint observation of the 2D PS and UV LFs.

Following \citet[][]{Park:2019}, we consider a limited selection of observed UV LFs at $z=6$ \citep{Bouwens:2017}, $z=7$ and 8 \citep{Bouwens:2015} and $z=10$ \citep{Oesch:2018}. This choice is motivated by limiting the systematic differences across the various groups within the literature and how each deals with their observational and statistical uncertainties. Ideally, to be truly robust (and conservative) one should average across all results in the literature to obtain a mean UV LF with a scatter encompassing all the differences across the various results. In future, we shall return to this while also extending our UV LF sampling to higher redshifts as observed by the JWST \citep[e.g.][]{Naidu:2022, Donnan:2022, Castellano:2022, Atek:2022, Harikane:2022a, Labbe:2022,Bouwens:2023,Willott:2023}. Importantly, when including UV LFs into our inference pipeline, we only consider UV magnitudes fainter that $M_{\rm UV} < -20$, for which it is argued that these are relatively dust-free \citep[see][]{Park:2019}.

In Figure~\ref{fig:wa-combined} we demonstrate the 1D and 2D marginalised posteriors for a mock 1000hr observation of the 2D PS in addition to UV LFs at $z=6, 7, 8$ and 10. Below the diagonal the orange (black) contours correspond to the 95th percentile joint marginalised posteriors when considering wedge removal with (without) UV LFs. Above the diagonal, we provide the equivalent following foreground avoidance with the magenta (purple) dashed contours denoting with (without) UV LFs. Finally, in Table~\ref{tab:results} we summarise the constraints and 68th percentile marginalised uncertainties.

In both cases, it is clear that the complimentary constraining power from the UV LFs improves the overall constraints on our astrophysical parameters using the 2D PS. In general, we find the amplitude of the improvements are larger for wedge avoidance relative to wedge removal. For example, we recover improvements of $\sim20$ and $\sim30$ per cent for $\alpha_{\ast}$ and $\alpha_{\rm esc}$ along with $\sim15$ per cent improvements for $M_{\rm turn}$. Whereas for wedge removal, at most we see improvements of $\sim10$ per cent for these same parameters. These relatively larger gains for foreground wedge avoidance following the inclusion of the UV LF information are due to the originally broader constraints and stronger degeneracies on the UV galaxy parameters, namely $\alpha_{\ast}$ and $f_{\rm esc,10}$. For the 2D PS with perfect foreground removal, since we have additional structural information on the EoR morphology through $k_{\perp}$, we are able to limit the degeneracy between $\alpha_{\ast}-f_{\rm esc,10}$ (see Figure~\ref{fig:wa}). Therefore, since this degeneracy is already reduced, the relative gains for the 2D PS with perfect foreground removal with UV LFs are also reduced.

Interestingly, once UV LFs are included with the 1D and 2D PS, the resultant 68th percentile uncertainties on our astrophysical parameters are reduced between the two PS. At most, we recover improvements of $\sim10$ per cent on the 2D PS + UV LFs relative to the 1D PS + UV LFs. This holds for either foreground mitigation strategy, with foreground removal still notably outperforming foreground avoidance. The origin of this stems from where the 2D PS gains its additional constraining power relative to the 1D PS. As highlighted earlier, the 2D PS is more sensitive to the EoR morphology as the structural information, $k_{\perp}$, is kept distinct from the redshift evolving component of the 21-cm signal ($k_{\parallel}$). This enables the 2D PS to improve over the 1D PS at constraining the EoR parameters, notably reducing the degeneracy between $\alpha_{\ast}$ and $f_{\rm esc,10}$. When including UV LFs, this serves a similar purpose, by adding additional information on $f_{\ast}$ to break the same degeneracy. Therefore, UV LFs add less unique information to the 2D PS than they do for the 1D PS. However, this behaviour is likely dependent on the underlying astrophysical model parameterisation. A model with additional parameters or more complex scalings with mass or redshift that are more sensitive to the EoR morphology would more significantly benefit from the 2D PS compared to the 1D PS, given how it better samples this information. Thus, in those instances the 2D PS + UV LFs would outperform the 1D PS.+ UV LFs as the UV LFs likely would add little additional information to more complex parameterisations.

\section{Conclusions} \label{sec:conclusion}

In recent years, simulation based inference (SBI) has begun to gain traction for performing Bayesian inference from the 21-cm signal to gain insights into the galaxies responsible for reionisation \citep[e.g.][Greig et al. in prep]{Zhao:2022,Zhao:2022b,Prelogovic:2023,Saxena:2023}. The significant advantage of SBI is that it applies machine learning principles to bypass the requirement to have an analytic expression to describe the likelihood function to accurately describe our 21-cm summary statistics. By removing this crucial bottleneck we are now able to rigorously explore more complex summary statistics than the simple, but extensively explored 1D spherically averaged power spectrum (1D PS). As a demonstration of the power of SBI, in this work we explore using the 2D cylindrically averaged PS (2D PS), which has previously been overlooked owing to the complexities in computing its likelihood.

For exploring the 2D PS we consider a mock 1000 hr observation of the 21-cm signal using the SKA. Throughout, we simulate the 21-cm signal using \cmfst{} \citep{Mesinger:2007p122,Mesinger:2011p1123,Murray:2020}, in particular the flexible UV galaxy parameterisation introduced in \citet{Park:2019}. As a result we have an eight parameter astrophysical model to describe the UV and X-ray properties of the first galaxies responsible for driving reionisation. Further, we consider two foreground mitigation strategies: (i) perfect foreground removal whereby we have access to the whole 2D information and (ii) foreground avoidance where we only use the pristine cosmological signal above the foreground contaminated wedge. Throughout, we perform SBI using marginal neural ratio estimation to learn the likelihood-to-evidence for performing parameter inference using \textsc{Swyft} \citep{Miller:2022}.

When considering perfect foreground removal, we find the 2D PS outperforms the 1D PS by reducing the 68th percentile uncertainties on individual parameters by up to $\sim30-40$ per cent. These relative improvements in the 2D PS over the 1D PS are consistent with recent predictions using the amplitude of the Fisher Information \citep{Prelogovic:2024}. Primarily, the most significant gains are in $M_{\rm turn}$ which effectively describes the minimum mass for star-forming galaxies along with $\alpha_{\ast}$ and $\alpha_{\rm esc}$ which describe the mass dependence of star-formation efficiency, $f_{\ast}$ and IGM escape fraction, $f_{\rm esc}$. These improvements are achieved due to the 2D PS cleanly separating the transverse information, $k_{\perp}$, from the redshift-evolving component of the signal, $k_{\parallel}$. In this way, we are more sensitive to the redshift evolution of the ionisation morphology allowing for improved constraints on the UV galaxy parameters. Unlike the 1D PS which combines and averages the anisotropic information into a single $k$ when spherically averaging.

Even when performing foreground avoidance, when we lose a large fraction of the 2D PS information relative to the case of perfect foreground removal, the 2D PS still outperforms the 1D PS. However, the relative boosts in performance are reduced, with only $20-30$ per cent improvements on our individual model parameters. Nevertheless, the largest gains remain for $M_{\rm turn}$, $\alpha_{\ast}$ and $\alpha_{\rm esc}$. This implies that despite the loss of a large fraction of information due to foreground contamination, distinguishing between the spatial ($k_{\perp}$) and frequency dependent ($k_{\parallel}$) Fourier modes still yields additional constraining power over the 1D PS for constraining the UV galaxy parameters during the EoR.

Comparing the two foreground mitigation strategies directly, we find foreground avoidance results in increased 68th percentile uncertainties of at worst $\sim2-3$ compared to foreground removal. In general, the largest increases are for the X-ray parameters, which are due to the growth of the foreground contaminated region towards larger redshifts, where the 21-cm signal is more sensitive to the X-ray contribution. However, we also see reductions at a similar level for $\alpha_{\rm esc}$ and $M_{\rm turn}$, owing to the loss of a significant fraction of spatial ($k_{\perp}$) information due to foreground wedge contamination. For the remainder, the 68th marginalised uncertainties increase by $\leq70$ per cent.

Finally, we also include independent astrophysical information by considering UV galaxy LFs at $z=6-10$. Doing so, we find improvements of $\sim10$ per cent primarily on $\alpha_{\ast}$, $\alpha_{\rm esc}$ and $M_{\rm turn}$ for foreground removal. For foreground avoidance, we find improvements of up to $\sim20-30$ per cent for these same parameters. Generally speaking, for the 1D PS the addition of UV LFs is to break the degeneracy between $f_{\ast}$ and $f_{\rm esc}$. However, for the 2D PS, as it is more sensitive to the EoR morphology through the distinct spatial information, the $f_{\ast}$-$f_{\rm esc}$ degeneracy is not nearly as strong. Therefore, the UV LFs have reduced benefit for foreground removal over foreground avoidance as we have additional 2D spatial information to reduce this otherwise strong degeneracy.

The power of SBI is that it enables the study of complex and non-Gaussian summary statistics of the 21-cm signal to be explored in the context of astrophysical parameter inference. Here, we have demonstrated the value of SBI with the first study of the more complex 2D PS. In future, to maximise the wealth of information expected to be available from the 21-cm signal we will explore alternative non-Gaussian statistics with SBI.

\section*{Acknowledgements}

Parts of this research were supported by the Australian Research Council Centre of Excellence for All Sky Astrophysics in 3 Dimensions (ASTRO 3D), through project number CE170100013. Y.S.T. acknowledges financial support from the Australian Research Council through DECRA Fellowship DE220101520. A.M. acknowledges support from the Ministry of Universities and Research (MUR) through the PRIN project ”Optimal inference from radio images of the epoch of reionization” as well as the PNRR project ”Centro Nazionale di Ricerca in High Performance Computing, Big Data e Quantum Computing”.

\section*{Data Availability}

The data underlying this article will be shared on reasonable request to the corresponding author.


\bibliographystyle{mnras}
\bibliography{Papers} 



\appendix

\section{Assessing Network Coverage} \label{sec:coverage}

\begin{figure*}
	\includegraphics[trim = 0.1cm 0.3cm 0cm 0cm, scale = 0.96]{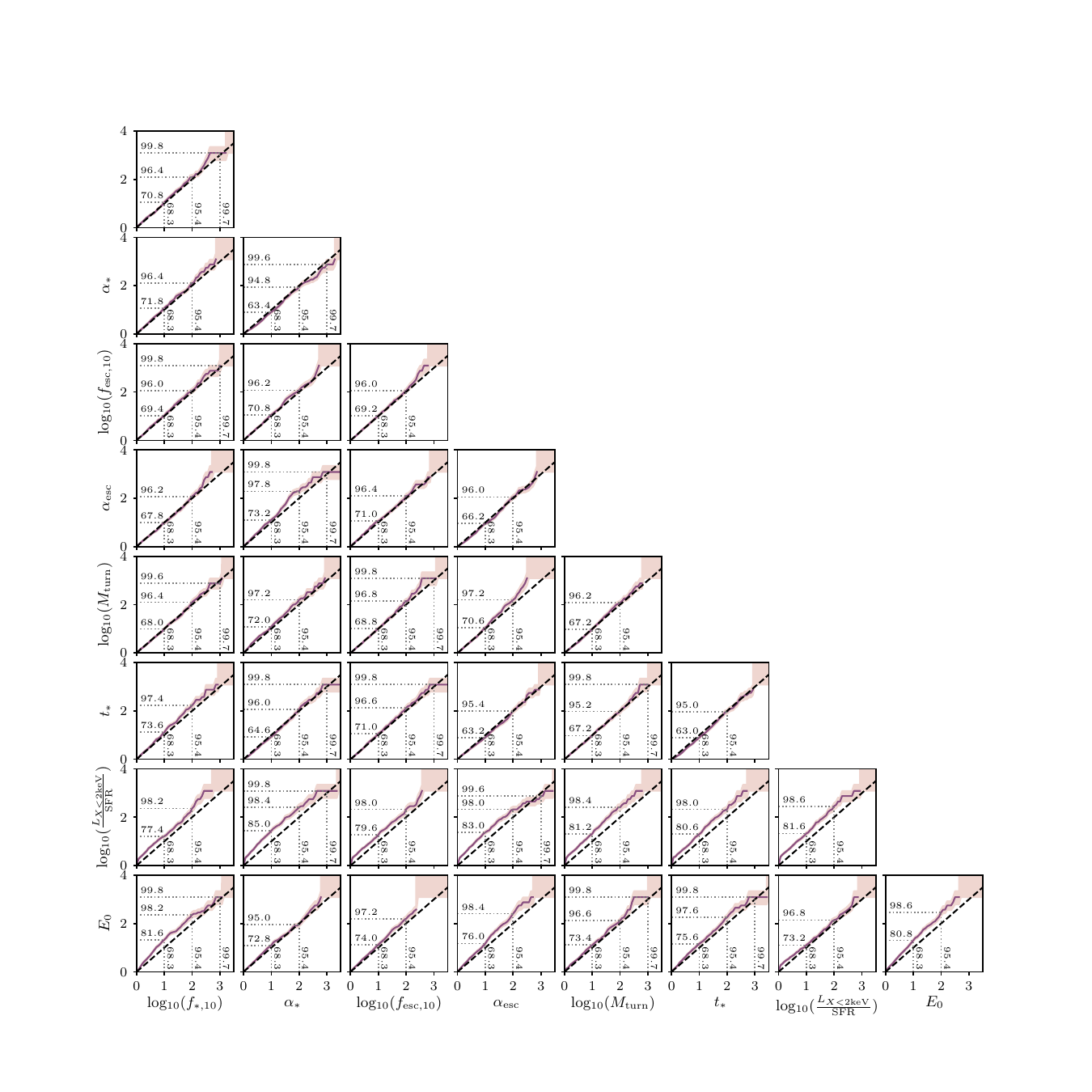}
    \caption{The empirical expected coverage probability of our trained MNRE network with \textsc{Swft} (vertical axis) as a function of the confidence level (horizontal axis). The purple line demonstrates the coverage of our network, with the goal of perfect coverage denoted by the diagonal black dashed line. The dotted lines indicate the coverage for the 68th, 95th and 99.7th percentiles whereas the shaded region corresponds to the Jeffrey's interval (see text for further details).}
    \label{fig:coverage}
\end{figure*}

One of the key defining features of SBI approaches such as MNRE is that once the network is trained they allow for the rapid recovery of the posteriors for any new realisation of the input data. In this case, we can perform parameter inference for a large number of mock observations drawn from within our prior range to determine how frequently they fall within their predicted posteriors. Measuring this frequency for a sufficiently large number of models enables the computation of the network coverage \citep[e.g.][]{Cole:2022}. This large number of direct posterior evaluations provides a much more robust quantity to indicate network convergence than those typically adopted by direct MCMC approaches \citep[e.g.][]{Betancourt:2019,Roy:2020}.

Following \citet{Cole:2022}, we define $\Theta_{\hat{p}(\hat{\boldsymbol{\theta}}|\boldsymbol{x}_{i})}(1-\alpha)$ to be a function which determines the $(1-\alpha)$ highest probability density region (HPDR) for our estimated posterior, $\hat{p}(\hat{\boldsymbol{\theta}}|\boldsymbol{x}_{i})$, given the input model-parameter pair, $\boldsymbol{x}_{i},\boldsymbol{\theta}^{\ast}_{i}$. To demonstrate, a 95 per cent HPDR would correspond to $\alpha=0.05$. For a set of $n$ independently drawn model-parameter pairs we can then determine the actual error rate, $1-\hat{\alpha}$ of the HPDR given our estimated posterior:
\begin{eqnarray}
1 - \hat{\alpha} = \frac{1}{n}\sum^{n}_{i=1} \mathds{1} \left[\boldsymbol{\theta}^{\ast}_{i} \in \Theta_{\hat{p}(\hat{\boldsymbol{\theta}}|\boldsymbol{x}_{i})}(1-\alpha) \right].
\end{eqnarray}
The quantities $\alpha$ ($\hat{\alpha}$) are re-defined in terms of a new variable, $z$, corresponding to the $1-\alpha/2$ ($1-\hat{\alpha}/2$) quantile of the standard normal distribution. By definition this implies the 1, 2, 3$\sigma$ regions correspond to $z=1,2,3$ with $1-\alpha = 0.6827, 0.9545, 0.9997$. The uncertainties on the error rate $\hat{\alpha}$ are determined by the Jeffreys interval \citep{Cole:2022}\footnote{Specifically, this interval is obtained from the 68.27 per cent central interval of a Beta distribution defined by the parameters ($n-k+1/2,k+1/2$) where $n$ is the total number of samples from the joint model and $k$ is the number of times the HPDR predicted by the network does not contain the true astrophysical parameters.}. In Figure~\ref{fig:coverage} we present the empirical expected coverage probability of our trained network as a function of confidence levels for all 1D and 2D marginalised posteriors. Optimal network performance is demonstrated by the black dashed curves. If the coverage probability resides above the black dashed line, the network coverage is deemed conservative (i.e. our actual error rate is lower than the theoretical error implying larger than expected posteriors uncertainties) whereas if it is below the diagonal it is considered over-confident. For the vast majority of our astrophysical parameters, our coverage probability is aligned or above the black dashed line indicating strong coverage performance.


\bsp	
\label{lastpage}
\end{document}